\documentclass[aip,amsmath,amssymb,sort&compress,numbers,reprint]{revtex4-2}

\usepackage[T1]{fontenc}
\usepackage[utf8]{inputenc}
\usepackage{float}

\usepackage{mathptmx}
    
\usepackage{setspace}
    
\usepackage{geometry}

\usepackage{adjustbox}
    
\usepackage{fancyhdr}
    
\usepackage[hidelinks]{hyperref}

\usepackage{titlesec}
    
\usepackage{acronym}
    
\usepackage[inkscapepath=svgsubdir]{svg}
    
\usepackage{enumitem}

\usepackage{tablefootnote}

\usepackage{multirow}
\usepackage{array}
\usepackage{makecell}
\usepackage{ltablex}

\usepackage{afterpage}

\usepackage{xcolor}
\usepackage{etoolbox}

\usepackage[section]{placeins}

\usepackage{tocloft}

\usepackage{textcomp}
\usepackage{gensymb}
\usepackage{pifont}
\newcommand{\mycomment}[1]{}
\usepackage{verbatim}
\date{}

\usepackage{amsfonts}
\usepackage{amsmath}
\usepackage{physics}

\usepackage{indentfirst}

\begin{document}

	\author{E. Filimonova}
	\email{evgeniia.filimonova@physik.uni-halle.de}
	\affiliation{Institute of Physics, Martin Luther University, 06099 Halle (Saale), Germany}

	\author{V. Ivanov}
	\affiliation{Institute of Physics, Martin Luther University, 06099 Halle (Saale), Germany}
	
	\author{T. Shakirov}
	\affiliation{Institute of Physics, Martin Luther University, 06099 Halle (Saale), Germany}

\title[Distinguishing noisy crystal symmetries in coarse-grained computer simulations]{Distinguishing noisy crystal symmetries in coarse-grained computer simulations: New procedures for noise reduction and lattice reconstruction}


\date{\today}

\begin{abstract}
We suggest new modification (we call it a noise reduction procedure) for Steinhardt parameters which are often used for detecting crystalline structures in computer simulation of solids and soft matter systems. We have also developed a new methodology how to reconstruct "ideal" lattice structure in the whole simulation box that would be most close to a real noisy crystalline symmetry, when it is defined locally and then averaged over the whole box. For this second procedure, which we call lattice reconstruction procedure, we have developed an algorithm for finding the lattice vectors from the values of Steinhardt parameters obtained after the noise reduction procedure. We apply noise to the classical crystalline structures ({\it sc, bcc, fcc, hcp}), and use both procedures to detect the crystalline structures in these classical but noisy systems. We demonstrate advantages of our procedures in comparison with existing methods and discuss their applicability limits. 
\end{abstract}

\maketitle
  \section{Introduction}
\label{sec:intro}
The model of hard spheres (HS) is widely used in theory and simulations of colloidal systems and it describes rather well experimental results on crystallization in colloidal systems \cite{Brown2018}. The model of tangent hard spheres (THS) is very popular for polymer systems. It describes flexible polymers if tangent hard spheres are freely jointed and semiflexible polymers if some bending potential on the angles between rigid bonds adjacent along the chain is applied. Phase transitions in such systems, in particular crystallization and liquid crystalline transition, are entropy driven. The Monte Carlo method is very convenient for computer simulation of such models and has been widely used for more than three decades \cite{dePablo1997, Loewen1997, Horbach2010, laso_ths1, laso_ths2, LasoPRL2009, LasoSoftMat2010, laso_ths3, LazoPhilMag2013}.

A very important (if not {\it the most important}) question in the studies of crystallization phenomena in soft matter systems is how to distinguish structural motifs, in particular ordered symmetries/morphologies, e.g., crystalline and liquid crystalline ones, on local and global scales in conformations which are usually quite noisy and often inhomogeneous, i.e., the local particle density and values of order parameters can fluctuate strongly, especially in the vicinity of phase transitions. 

Usually, the radial distribution function (RDF) and static structure factor are calculated first. For example, for freely jointed hard-sphere chains one can see more pronounced maxima in the solid phase in comparison to the liquid one \cite{rdf-FJTHS}, but one cannot reliably distinguish different crystalline symmetries by data on RDF.
Another parameters and methods of analysis for crystals are Voronoi tesselation \cite{voro, voro_manual}, common neighbor analysis \cite{cna, cna2, acna, icna}, nematic order parameter, chain segments local alignment parameter \cite{schilling, kos}.

Bond orientational order parameters or Steinhardt parameters \cite{steinhardt_main} have been suggested long ago to distinguish crystals with different symmetries. However, these parameters work well only for ideal crystals or for crystals with very weak fluctuations in the particle positions, but they need modifications for typical cases in real soft matter systems, and such modifications have been suggested \cite{dellago1, Mecke2013, eslami, Haeberle2019}. 

The characteristic crystallographic element (CCE) norm as a powerful descriptor of local structure in atomistic and particulate systems was introduced in Ref. \onlinecite{LasoCCE2009}. The CCE-norm is sensitive both to radial and orientational deviations from perfect local order. Recently, the CCE-norm has been revised and extended for reliable identification of local structure in 2d and 3d atomic systems \cite{LasoCryst2020}.
The local and global order in dense packings of semi-flexible chains consisting of tangent hard spheres has been studied in Ref. \onlinecite{LasoPolym2023-1}. It was analyzed how the packing density and chain stiffness influence the self-organization of chains at the local and global levels. The local order corresponds to crystallinity, and it was  quantified by the CCE-norm descriptor \cite{LasoPolym2021}, while the global order was computed through the scalar orientational order parameter \cite{LasoPolym2023-1}.

Recently, we have started investigation of surface phenomena in crystallization of semiflexible tangent hard spheres' chains by means of flat histogram Monte Carlo simulations \cite{samc_paul_janke}. We use the same model for which crystallization in the bulk has been studied before \cite{tim_model1}, but consider melts at different substrates modelled by either purely repulsive walls or by attracting walls with various depth of well potential. In our simulations we use all above mentioned parameters, except CCE-norm.  
However, we have found that all these parameters, including modifications of Steinhardt parameters, are still not entirely accurate for the systems which we get in our simulations. Therefore, we have suggested another new modification for Steinhardt parameters and called it a noise reduction procedure. We have also developed a novel methodology how to reconstruct "ideal" lattice structure in the whole simulation box that would be most close to a real local noisy crystalline symmetry when averaged over the whole box. For this second procedure, which we call lattice reconstruction procedure, we have developed an algorithm for finding the lattice vectors from the values of Steinhardt parameters obtained after the noise reduction procedure. We apply noise to the classical crystalline structures ({\it sc, bcc, fcc, hcp}), and afterwards use both procedures to determine the crystalline structures in classical but noisy systems. 

Our paper is organized as follows. In Section \ref{sec:model} we describe our model which we have used in our simulations of semiflexible chains of tangent hard spheres and present some examples of conformations which motivated us to developed the new modifications for structural analysis. We present here also our estimates of the noise amplitude in those systems, which we will use to apply noise to classical crystalline symmetries, It should be emphasized here, that our results on phase transitions in semiflexible THS systems will be published in a separate paper, and here we concentrate only on the methodological developments, which are (in our opinion) quite important for the field of computer simulations of soft matter systems, in particular for polymers, colloids and liquid crystals. In Section \ref{sec:structure_analysis} we present our results on structure analysis in noisy {\it sc, bcc, fcc, hcp} lattices by previously suggested methods and demonstrate the problems with correct detecting of the symmetries, Then, in Section \ref{subsec:noise_reduction} we present our new noise reduction procedure and discuss the results of structure analysis by this method. A Section on our new lattice reconstruction procedure will be added soon. Section \ref{sec:concl} contains our conclusions. 
  \section{Model}
\label{sec:model}

Flexible tangent hard-sphere chain model used in this work is well established in works devoted to the study of crystallization in short chain melts \cite{tim_model1, tim_model2}. 
The hard-sphere-type interactions of non-bonded beads:

\begin{equation}\label{nonbond}
U_{nb} (r)=
\begin{cases}
\infty, & r \leq \sigma ,\\
0, & r > \sigma ,
\end{cases}\end{equation}

\noindent where $r$ is the distance between the centers of the two beads. 
This interaction does not give a numerical contribution to the energy, but imposes constraints on the available configuration space. 
As a result, possible angles between the bonds of neighboring spheres $\theta$ in the chain cannot exceed $120^o$.
The lengths of chains in all studied systems are $N=10$ beads. 

\begin{figure}[h]
\begin{center}
\includegraphics[width=0.7\linewidth]{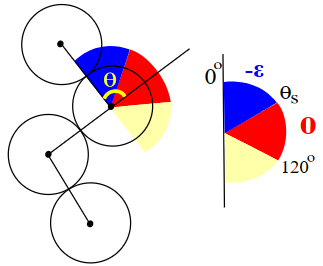} 
\end{center}
    \caption{illustration of stiffness potential}
    \label{fig:model1}
\end{figure}

Stiffness potential of the chains: 

\begin{equation}\label{stiffnes}
U_{\theta} (\theta)=
\begin{cases}
- \epsilon, & \theta \leq \theta_s ,\\
0, & \theta_s < \theta < 120^o  , \\
\infty, & 120^o \leq \theta ,
\end{cases}\end{equation}

The ground-state stiffness energy for systems of
$N_c$ chains of length $N$ is defined as $E^s_{min} = -N_c(N - 2) \epsilon$, while the maximum energy equal to $E^s_{max} = 0$. We consider several boxes with constant volume fraction $\phi\approx0.496$. The fixed value of $\cos(\theta_s)=0.9$ used in this study corresponds to  $ \theta_s\approx 26^o$, which at a volume fraction $\phi\approx0.496$ makes it possible to observe an ordered structure. 
During the simulation, the size of the box ($L_x$, $L_y$, and $L_z$) does not change.
Energies and temperatures are measured in units of well depth $\epsilon=1$, while all lengths are measured in units of solid sphere diameter $\sigma=1$.

\subsection*{System in the absence of walls}
We simulate the system that we will use as a reference system of the polymer in the unconstrained bulk. Since in small size systems, the finite size effects of the system have a great impact, we use 2 simulating boxes: with sizes  $L_x = L_y = 20$ and $L_z = 19$ $(N_c = 720)$ and  $L_x = L_y = 20$ and $L_z = 39$ $(N_c = 1440)$. This ratio of the number of particles and the size of the box provides the desired volume fraction:

\begin{equation}\label{vol_fraction}
\phi=\frac{NN_c}{L_x L_y L_z}\frac{\pi\sigma^3}{6}\approx0.496
\end{equation}

Since there are no constraints and there are periodic boundary conditions in all directions, the total energy $E$ is composed only of the stiffness energy $E^s$ of the chains:
\begin{equation}\label{en_st}
E = E^s = \sum U_{\theta}.
\end{equation}

\subsection*{System with two purely repulsive walls}
In the case of uniformly repulsive walls, a potential $U_{rep}$ is applied between the bead and the wall :
\begin{equation}\label{repulsion}
U_{rep} (z)=
\begin{cases}
\infty, & |z| \; \geq \; (L_z - \sigma) / 2  ,\\
0, & |z| \; < \; (L_z - \sigma) / 2,
\end{cases}\end{equation}

\noindent where $z$ represents a coordinate of a bead along $z$ axis. 
In fact, this potential hinders chain units from passing through the wall and does not contribute numerically to the total energy value. That is, the contribution to the energy is determined by the same summand (Eq. \ref{en_st}) as in the case of the unconstrained system. 
As the centres of the beads cannot come closer to the surface than their own radius of $\sigma/2$, we increase the size of the box along the $z$ axis to the value of $\sigma$ in order to ensure that we preserve the bulk volume fraction of the polymer, as in the system without walls $(\phi \approx 0.496)$.
Therefore, in the presence of walls, the dimensions are $L_x = L_y = L_z = 20$ $(N_c = 720)$ and $L_x = L_y = 20$ and $L_z = 40$ $(N_c = 1440)$. 
Thus, two parallel walls are positioned on the planes at $z=\pm 10$ and $z=\pm 20$ respectively. Along the $x$ and $y$ axes, periodic boundary conditions are implemented.

\subsection*{System with purely repulsive and attractive walls}
We also examine two scenarios involving a small system ( $L_x = L_y = L_z = 20$, $N_c = 720$ ) where one wall exhibits repulsion and is conserved at $(z = -10)$, while the second wall has an attracting potential at $(z = 10)$: 

\begin{equation}\label{attraction1}
U_{at_1} (z)=
\begin{cases}

- \epsilon, &L_z / 2 - \sigma  \;  \leq \; z \; < \; (L_z - \sigma) / 2 ,\\
0, &- (L_z - \sigma) / 2 \; < \; z \; < \; L_z / 2 - \sigma,\\
\infty, &|z| \; \geq \; (L_z - \sigma) / 2

\end{cases}\end{equation}

\begin{equation}\label{attraction2}
U_{at_2} (z)=
\begin{cases}

- 4 \epsilon, &L_z / 2 - \sigma  \;  \leq \; z \; < \; (L_z - \sigma) / 2  ,\\
0, &- (L_z - \sigma) / 2 \; < \; z \; < \; L_z/ 2 - \sigma,\\
\infty, &|z| \; \geq \; (L_z - \sigma) / 2 

\end{cases}\end{equation}

  \section{Structure analysis}
\label{sec:structure_analysis}

In general, a crystal means a solid body having a three-dimensional long-range translational order. 
The arrangement of atoms in a crystal is characterized by its unit cell containing one or more atoms in a certain spatial arrangement. 
The crystal structure of the substance is reproduced by periodic translation of the unit cell. 
In this study, we do not pretend to determine the exact class of the crystal, since this would represent a separate in-depth research in the field of crystallography. 
In addition, in polymer systems, we can only talk about local translational ordering and the formation of a set of crystallites. 
However, we claim to be able to detect the coexistence of various crystal symmetries.
Despite the fact that a significant number of methods for analyzing crystal structures have been developed, this task is still the cornerstone of computer simulation research. In this chapter, we will look at the most common methods of analysis applied to frequently considered systems such as simple cubic (sc), body-centered cubic (bcc), face-centered cubic (fcc), hexagonal close-packed (hcp), as well as to our system and propose another method of analysis that we consider the most successful in this study.
  \subsection{Noise estimation}
\label{subsec:noise_estimation}

Before starting to analyze the crystal structure, it is necessary to pay attention to the fact that one cannot expect to obtain an ideal crystal within the framework of the model used.
In the studied systems, noise will be observed due to the breadth of the potential.
Since the condition for adding the value $-1$ is an angle ranging from 0 to 26 degrees, then the lowest energy of $E=-5760$ in a small system will be given by both, as well as fully elongated chains with all angles between the chains equal to 0, and elongated chains with all angles close to 26 from below. 
Thus, the width of the potential is the cause of noise in the resulting crystal structures. 
This paragraph will describe the procedure for evaluating noise in our system.

\begin{figure}[ht]
\begin{minipage}[ht]{0.47\linewidth}
\center{\includegraphics[width=1\linewidth]{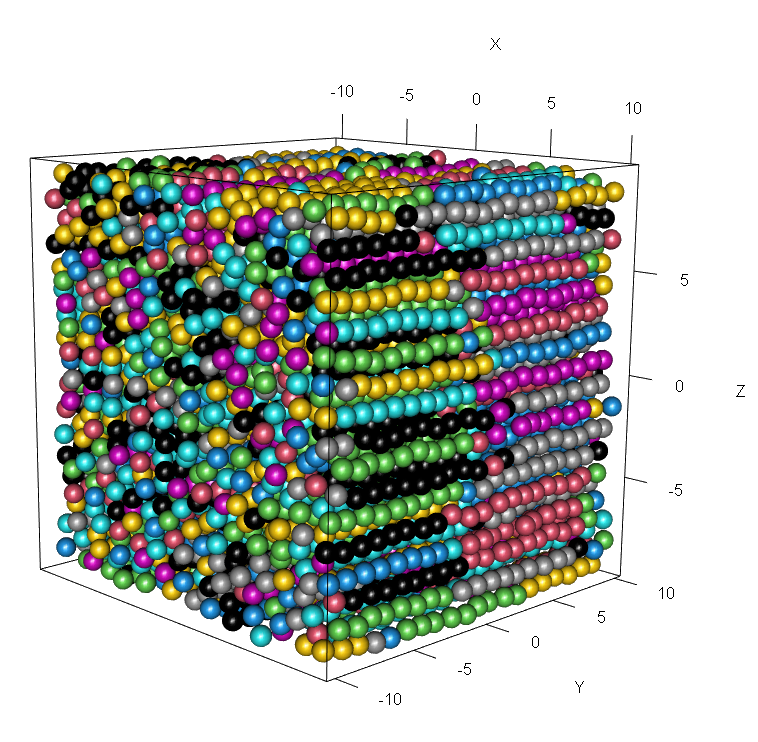}} \\ (a)
\end{minipage}
\hfill
\begin{minipage}[ht]{0.47\linewidth}
\center{\includegraphics[width=1\linewidth]{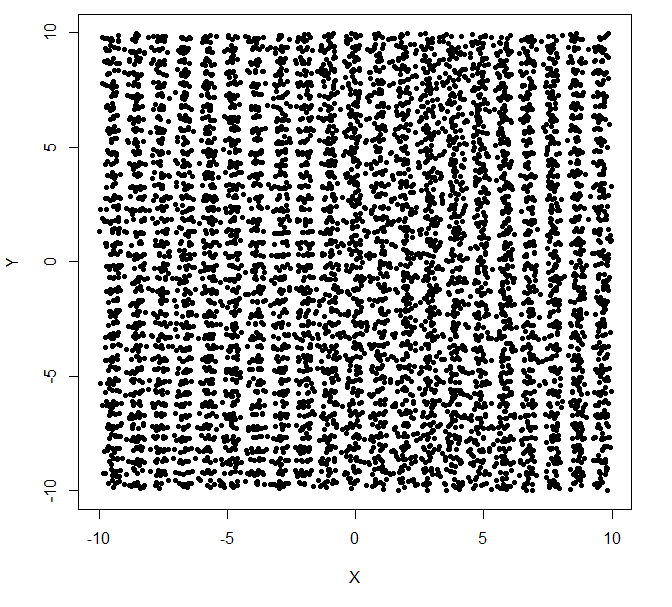}} \\(b)
\end{minipage}
\vfill
\begin{minipage}[ht]{0.47\linewidth}
\center{\includegraphics[width=1\linewidth]{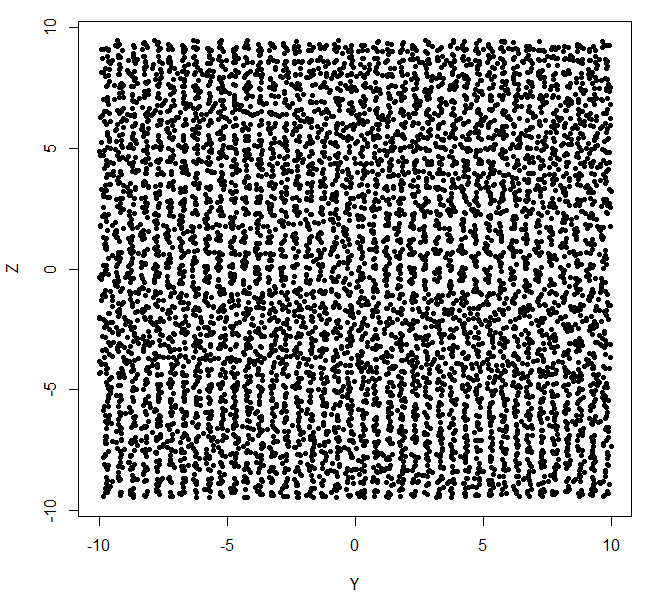}}  \\(c)
\end{minipage}
\hfill
\begin{minipage}[ht]{0.47\linewidth}
\center{\includegraphics[width=1\linewidth]{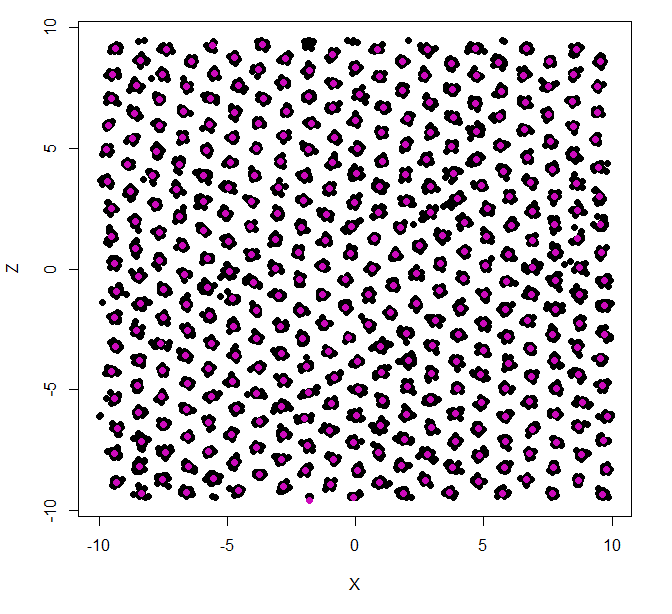}}  \\ (d)
\end{minipage}

    \caption{a snapshot of the system at energy E=-5727: (a) 3d view. The colors correspond to different chains; (b), (c), (d) the dots denote the centers of mass of each sphere in the projection on the plane $XY$, $YZ$, $XZ$ respectively. Purple points on (d) denote centers of groups.}
    \label{fig:bulk5727}
\end{figure}

Let's choose a system with a sufficiently low energy. For example, consider a system with energy $E=-5727$ (Fig.\ref{fig:bulk5727}), for other energies and systems, estimates give a similar result. 

\begin{figure}[ht]
    \center{
    \begin{minipage}[ht]{0.43\linewidth}
\center{\includegraphics[width=1\linewidth]{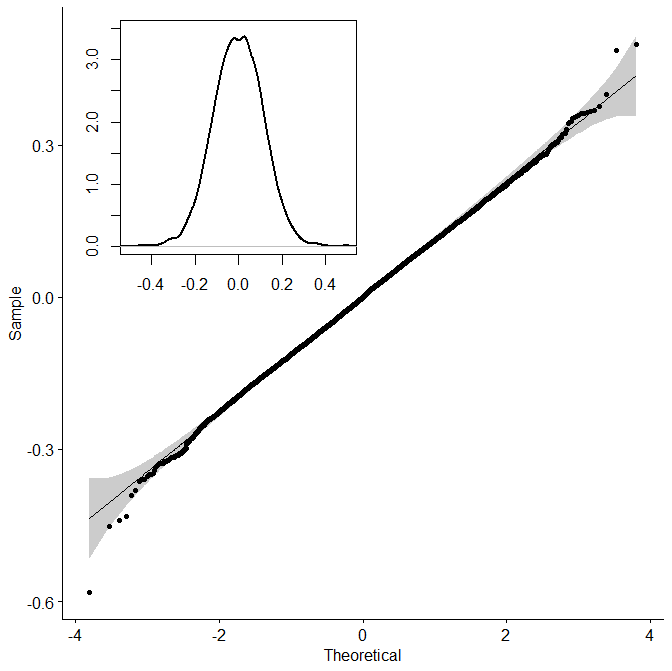} \\ (a)}
\end{minipage}
\begin{minipage}[ht]{0.43\linewidth}
\center{\includegraphics[width=1\linewidth]{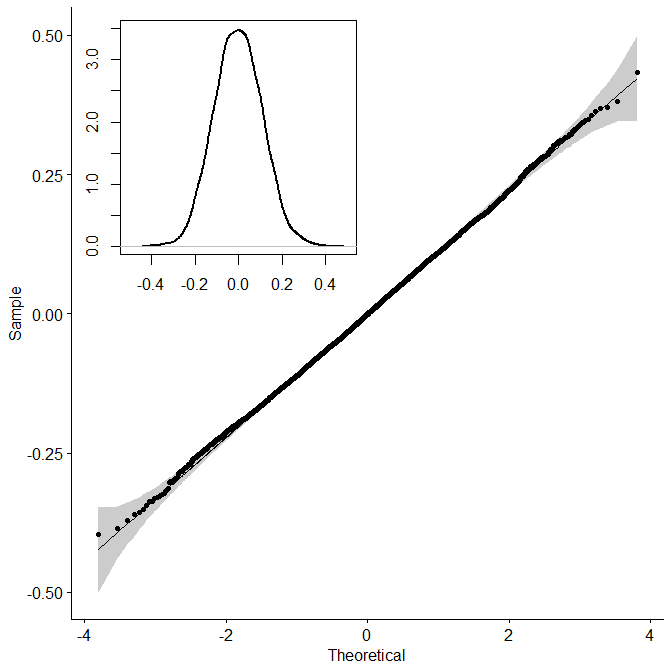} \\ (b)}
\end{minipage}
    \caption{main: quantile-quantile plot for deviation from the centers of the groups in $x$ direction (a) and $z$ direction (b); inserts: deviation from the centers of the groups in $z$ direction (a) and $z$ direction (b).}
    \label{fig:normtest}
    }
\end{figure}

While in the $XY$ and $YZ$ planes one can notice a variety of patterned structures of particles, in the $XZ$ plane one can observe uniting into groups. 
In the above case, the organization of projections into groups in this plane is explained by the elongation of the chains along the $y$ axis (Fig.\ref{fig:bulk5727} a). Thus, the center of the groups has the meaning of the atom of the crystal lattice, and the size of the group characterizes the noise in the crystal.
At this stage, clustering was carried out using the k-means method \cite{kmeans} (with an input number of 360 groups) and a method similar to k-means, however, not the number of groups was set as an input parameter, but the initial roughened size of the groups ($R=0.8$). Both methods give the same result for groups positions.
As can be seen from the inserts in Fig. \ref{fig:normtest}, the distribution of deviations of points from the centers of groups resembles the Gaussian distribution. 
One can use quantile-quantile plot (QQ plot) to check normality visually. 
QQ plot draw a correlation between the sample and the normal distribution. 
In a QQ plot, each observation is displayed as a single point. 
If the data correspond to the norm, then the dots should form a straight line. 
Thus, in our case one can talk about the normality of the distribution. 
In addition to the ability to visually assess whether the distribution belongs to the normal distribution, this method allows estimating the median value of the distribution and the standard deviation. 
The value on the $y$ axis at $x=0$ corresponds to the median. 
We see that it is equal to zero; this correctly reflects our construction: we translated the center of each group to the origin before counting the deviation of the position of the particle of each group from its center. 
The tangent of the slope of the approximating lines corresponds to the standard deviation of the distribution. 
From the resulting plots we can extract $\sqrt{\sigma^2} \approx 0.112$. 
This value characterizes the noise in our system. 

Another way to estimate the parameter $\sqrt{\sigma^2}$, assuming that the distribution of deviations of particles from the centers of groups $D(x)$ is normal, is to construct a $\ln{D(x)}$ :

$$ 
D(x) \sim \exp \Big\lbrace- \frac{(x-x_0)^2}{2\sigma^2} \Big\rbrace $$
$$
\ln D(x) = const_1 - \frac{(x-x_0)^2}{2\sigma^2}$$
Since we assume that the centers of all groups have already been translated to the origin, then $x_0 = 0$.
$$\frac{\partial\ln D(x)}{\partial x} = - \frac{x}{\sigma^2}
$$
As can be calculated from the coefficients obtained in the linear model $\sqrt{\sigma^2}$ is in both cases: 
\begin{equation}\label{Sigma}
\sqrt{\sigma^2} \approx 0.112,\;\; (\sigma^2 \approx 0.0125 ).    
\end{equation}

Taking into consideration that this is an estimate for one group along one direction, for a pair of noisy groups in three-dimensional space we get $6\sigma^2 \approx 0.075 $.
We will use this estimation in the following parts of the work. 

\begin{figure}[ht]
\center{
    \begin{minipage}[ht]{0.40\linewidth}
\center{\includegraphics[width=1\linewidth]{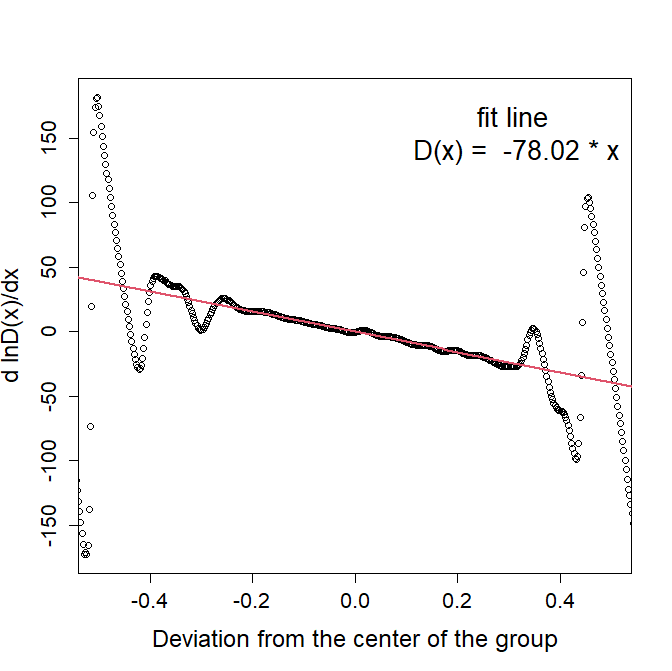} \\ (a)}
\end{minipage}
\begin{minipage}[ht]{0.40\linewidth}
\center{\includegraphics[width=1\linewidth]{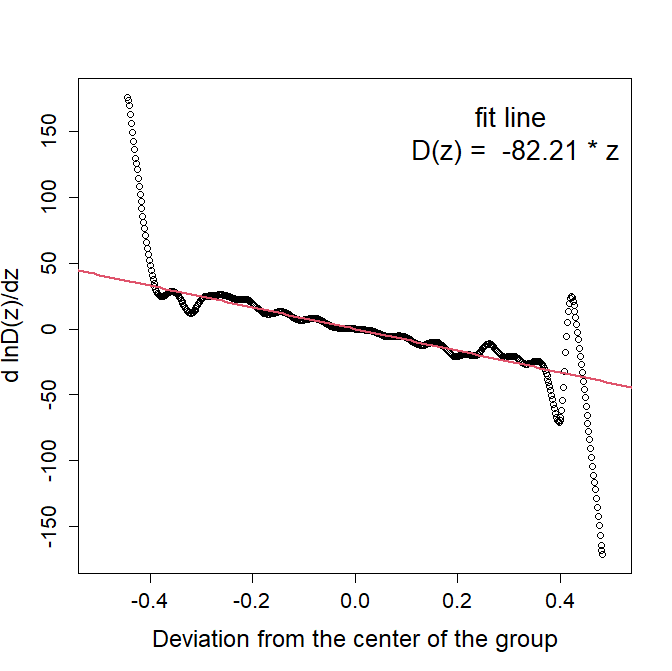} \\ (b)}
\end{minipage}
    \caption{noise estimation in $x$ direction (a) and $z$ direction (b).}
    \label{fig:noise_est}
    }
\end{figure}

  \subsection{Test structures preparation}
\label{subsec:prep_test}

In order to make sure that methods used are reliable, we will perform tests on well-known structures: $sc$, $bcc$, $fcc$, $hcp$.
For these structures primitive vectors are known, by which it is possible to reproduce the crystal lattice. 
In order that the results obtained for the test lattices could be compared with the structures studied in this work, two conditions were taken into account: 
{\sf(a)} the minimum distance between the lattice atoms of the perfect test structures should be equal to the length of the rigid bond in the polymer model under study, i.e. $d=1$; 
{\sf(b)} the atomic packing factor ($APF$) of the test structures should coincide with the volume fraction of the polymer in the system under study ($APF = \phi = 0.496$), since this parameter is the same for all the systems under study and does not change during simulating process.

Since the dependence of various parameters on noise $\sqrt{\sigma^2}$ is investigated and special attention is paid to the value estimated in the section \ref{subsec:noise_estimation} ($\sqrt{\sigma^2}=0.11$), here we will also describe the procedure for introducing noise $\sqrt{\sigma^2}$ into the test structures taking into account the excluded volume.

{\sf{\it Example}}

As an example, let us consider the construction of $bcc$ structure on the basis of which comparisons of the investigated parameters will be made. 
The construction of other test structures is carried out in full analogy.

Primitive lattice vectors for $bcc$: 
\begin{eqnarray}
\vb{a} = -\frac{a}{2}\vb{e_1} + \frac{a}{2}\vb{e_2} + \frac{a}{2}\vb{e_3}, \nonumber \\
\vb{b} = \frac{a}{2}\vb{e_1} - \frac{a}{2}\vb{e_2} + \frac{a}{2}\vb{e_3}, \\
\vb{c} = \frac{a}{2}\vb{e_1} + \frac{a}{2}\vb{e_2} - \frac{a}{2}\vb{e_3},  \nonumber
\end{eqnarray}

\noindent where $a$ is the lattice parameter, or the side of the cube, related to the distance between the nearest lattice atoms $d$ by the relation
\begin{equation}\label{l1}
a=\frac{2}{\sqrt{3}}d. 
\end{equation}
\noindent Usually, to calculate the  $APF$ one assume that $d=2r$, where $r$ is the radius of the particle.
However, to adjust the $APF$, we fix $d=1$, according to assumption {\sf(a)} and change the particle radius $r$. $APF_{bcc}$ is expressed in terms of the volume of a single particle $V_1$, the number of particles in a unit cell $N_{bcc}$ and the volume of a unit cell $V_0$:

\begin{equation}
    APF_{bcc}=\frac{V_1 \cdot  N_{bcc}}{V_0} = \frac{\frac{4}{3} \pi r^3 \cdot 2}{a^3}.
\end{equation}

After using the condition Eq. \ref{l1}, and the requirements {\sf(a)} and {\sf(b)} that $APF_{bcc} = \phi = 0.496$ from Eq. \ref{vol_fraction}, one obtains:

\begin{equation}\label{l3}
    r = \Bigl({\frac{\phi}{\pi\sqrt{3}}}\Bigr)^{\frac{1}{3}} \approx 0.45.
\end{equation}

This value is particularly important in the procedure for the noise introducing to the hard sphere system.

{\sf{\it Procedure for introducing noise}}

We assume that the noise along any direction in our system has a Gaussian distribution. 

Therefore, we choose orthogonal vectors $\vb{e_1}, \vb{e_2}, \vb{e_3}$, and we will drive noise along each of these directions. 

\underline{Step 1}

At the first step in the constructed ideal lattice, each particle acquires a normally distributed Gaussian displacement 

\begin{equation}
\Delta x_1 = \mathcal{N} (0,\sigma^2)
\end{equation}
Since at the initial stage we did not control the overlap of particles, now we need to eliminate it.  

\underline{Step 2}

We are looking for a pair of points $i_0$, $j_0$ that form the minimum pair distance $d_{min}$ in the resulting noisy system. 

\underline{Step 3}

Randomly select one particle $i\;^0$ from $i_0$ and $j_0$. With a probability of $0.1\%$, the particle $i\;^0$ is randomly selected from all the particles of the system. For the particle $i\;^0$, the displacement introduced earlier is replaced by:

\begin{equation}\label{l2}
\Delta x_2 \longrightarrow \alpha \Delta x_2 + \mathcal{N} (0,\sigma_1^2)
\end{equation}

\noindent where $\alpha \in [0;1]$. 
According to the properties of the normal distribution, the distribution $\Delta x_2$ will be distributed normally with a width of $\sqrt{\sigma^2}$ if one sets:

\begin{equation}
\sigma_1^2 = (1-\alpha^2)\sigma^2.
\end{equation}

An attempt to change the offset (Eq. \ref{l2}) is accepted if the paired distances to the particle $i\;^0$ taking into account this step are not less than $d_{min}$, otherwise $i\;^0$ retains its previous offset $\Delta x_1$. 

\underline{Step 4}

If a new displacement $\Delta x_2$ is accepted, a new $d_{min}$ and particles $i_0$, $j_0$ are searched.

\underline{Step 5}

Repeat Step 2 - Step 5 until $d_{min} < 2 r$, where $r$ was evaluated in Eq. \ref{l3}.

\underline{Remark}

After completing the procedure, the resulting structures were checked for the normality of the distribution of the position of the particles relative to the noiseless positions, similar to how it was done in the section \ref{subsec:noise_estimation}.
The test showed that this method works well in our task for small values of $(\sqrt{\sigma^2} < 0.07)$. 
When $\sqrt{\sigma^2}> 0.07$, the dependence of the observed noise on the $\sqrt{\sigma^2}$ value we introduce during the procedure is no longer linear. 
For example, for $hcp$, the observed noise becomes comparable to that observed in our system (Eq. \ref{Sigma}) at induced $\sqrt{\sigma^2} = 0.16$. 
Later in the text, speaking about noise and using the notation $\sigma$, we will use the actually observed values obtained during this check.

  \subsection{Radial distribution function}
\label{subsec:rdf}

Let's start the study of structures by calculating the radial distribution function (RDF). 
Let $G(r)$ be the probability density of the presence of a particle at a distance $r$ from a given particle. 
Then the probability of the presence of a particle at a distance $r$ from this particle is defined as $G(r)dr$. 
Then the radial distribution function $g_2(r)$ is calculated as:

\begin{equation}
    g_2(r) = \frac{G(r)dr}{4\pi r^2 dr} V,
\end{equation}
where $V$ is the volume under study. 

\begin{figure}[h]
\center{
    \begin{minipage}[h]{0.47\linewidth}
\center{\includegraphics[width=1\linewidth]{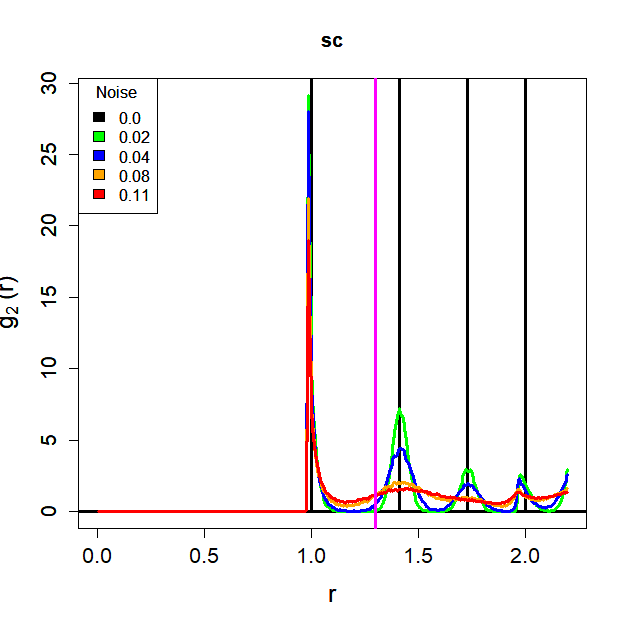} \\ (a)}
\end{minipage}
\begin{minipage}[h]{0.47\linewidth}
\center{\includegraphics[width=1\linewidth]{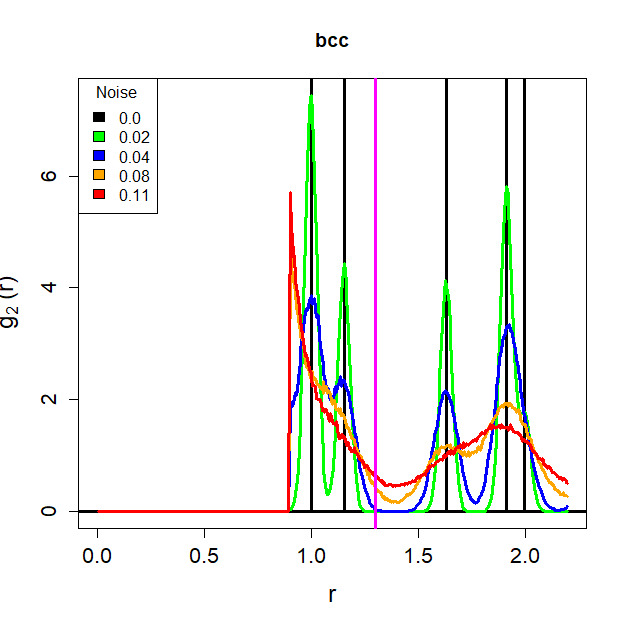} \\ (b)}
\end{minipage}
    \vfill
    \begin{minipage}[h]{0.47\linewidth}
\center{\includegraphics[width=1\linewidth]{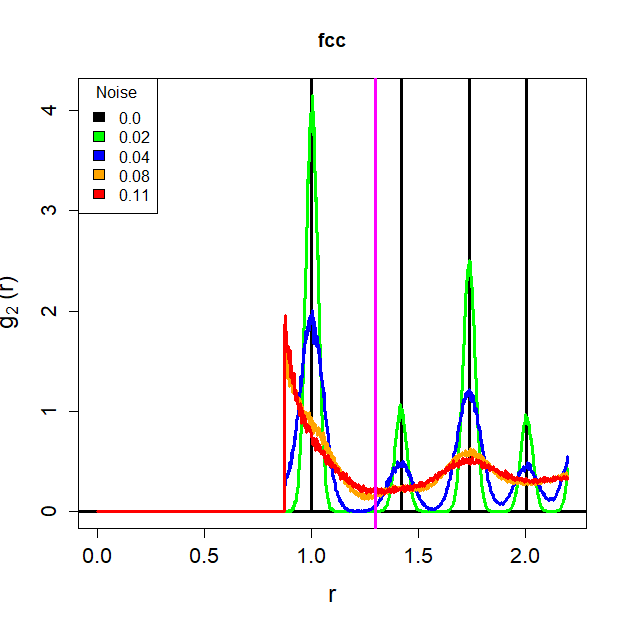} \\ (c)}
\end{minipage}
\begin{minipage}[h]{0.47\linewidth}
\center{\includegraphics[width=1\linewidth]{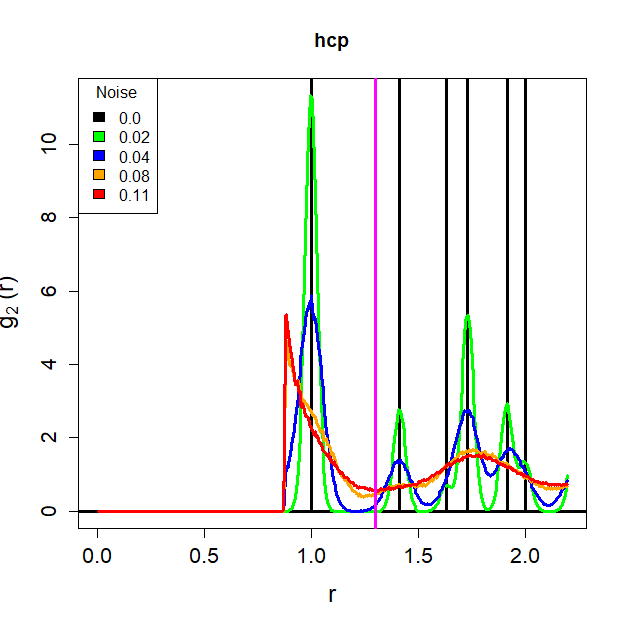} \\ (d)}
\end{minipage}
    \caption{radial distribution function for different noises for $sc$ (a), $bcc$ (b), $fcc$ (c), $hcp$ (d). 
    \label{fig:rdf_noise}
 }}
\end{figure}

As it was established in the section \ref{subsec:noise_estimation}, the formation of a perfect crystal cannot be expected in the structure under the study.
Noise can contribute to the displacement and blurring of the radial distribution function.
To study the behavior of the $g_2(r)$ depending on noise, $sc$, $bcc$, $fcc$, $hcp$ structures were selected as test structures.
Gaussian noise $\sqrt{\sigma^2}$ is introduced into the initially perfect crystal lattice in each of the three directions, as it was described in the section \ref{subsec:prep_test}. 
The red curves (Fig.\ref{fig:rdf_noise}) correspond to noise similar to the system under study, since noise $\sqrt{\sigma^2}\approx0.11$ is introduced in this case. It can be seen from the data obtained that the number of peaks on the RDF decreases and the determination of the symmetry of the structure becomes impossible. 
The position of the first maxima shifts to the left on all structures with increasing noise. For small $\sqrt{\sigma^2}$ values, the maxima of all structures are located at the point $r=1.0$, which corresponds to the method of constructing lattices described in section \ref{subsec:prep_test}. 
This value corresponds to the minimum distance between the lattice nodes. With a sufficiently large $\sqrt{\sigma^2}$, the peak corresponds to the minimum possible distance that the lattice atoms can approach, that is, the diameter of solid spheres. 

\begin{figure}[h]
\center{
\center{\includegraphics[width=0.6\linewidth]{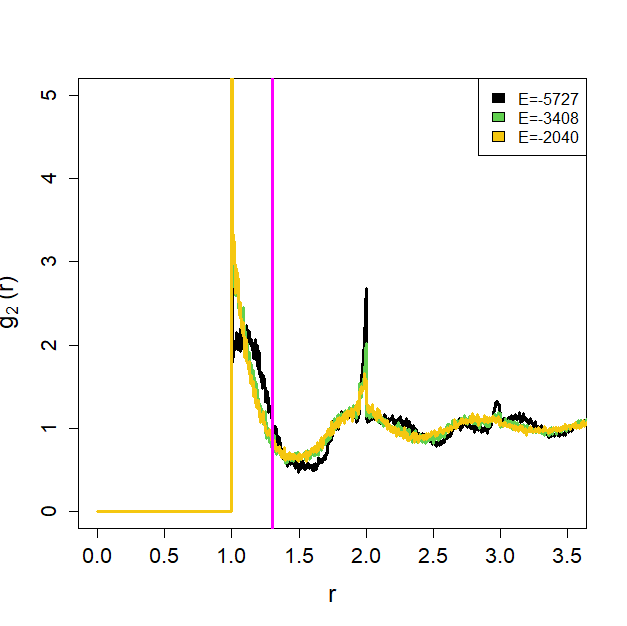} }

    \caption{radial distribution function for studied structures. The magenta line corresponds to the chosen cutoff radius $R_c=1.3$ when further calculating the order parameters.
    \label{fig:rdf}
 }}
\end{figure}

On the graphs of $g_2(r)$ (Fig.\ref{fig:rdf}) for the system under study (at energies $E =-2040$ and $E=-5727$, which corresponds to a melt and an ordered structure respectively), peaks at $r=1$ and $r=2$ can be observed. These maxima correspond to neighboring particles along the chain in our model. 

The calculation of the RDF is a necessary step for calculating the Steinhardt parameters. 
In general, Steinhardt and co-authors \cite{steinhardt_main} recommend using particles that fall into the sphere of the cutoff radius $R_c=1.2r_0$ as the nearest neighbors, where $r_0$ is the position of the first peak of the $g_2(r)$. 
Such a choice should ensure that all particles in the first coordination sphere are taken into account. 
However, since the noise estimation in our system is $6\sigma^2=0.27$, it was decided to use the $R_c=1.3r_0$. Since $r_0=1$ coincides with the diameter of the hard spheres in further text we will immediately write $R_c=1.3$. Our experience has shown that this value is optimal for calculating parameters. 
If we consider the distributions for the model under study (Fig.\ref{fig:rdf}), one can see that the selected value of the $R_c$ is to the left of both minima. 
As this value is close to the minima of both our structure and the test ones, this choice is the most optimal, since it is most likely to capture all particles from the first coordination sphere ($R_c=1.3$ magenta in the Fig. \ref{fig:rdf_noise}).

  \subsection{Local bond order parameters}
\label{subsec:local}
\subsubsection{P. J. Steinhardt, R. Nelson and M. Ronchetti parameters}
\label{subsubsec:steinhardt}

One of the most discussed problems of simulation of the crystallization process is how to attribute a particle to a liquid or crystal and also determine the type of crystal. 
Recently, a widely used method of distinguishing the type of particles is local bond order parameters, also known as Steinhardt parameters \cite{steinhardt_main}. 
To calculate these parameters, it is not necessary to have a reference structure for comparison with the studied one, as is required in, for example, common neighbor analysis \cite{cna}. 
Another convenience of using this method is the absence of reference to the coordinate system, since the calculation of these parameters is based on spherical harmonics. 

Local bond order parameter of particle $i$ is defined as:

\begin{equation}\label{stein_eq}
    q_{l}(i) = \sqrt{ \frac{4\pi}{2l+1}\sum_{m=-l}^l \abs{q_{lm}(i)}^2 },
\end{equation}

\noindent where $l$ is the order of parameter, $q_{lm}$ - complex vector which is defined as

\begin{equation}
    q_{lm}(i) = \frac{1}{N_b(i)} \sum_{j=1}^{N_b(i)} Y_{lm}(\vb{r}_{ij}).
\end{equation}

\noindent Here, $N_b(i)$ is the number of nearest neighbors of the particle $i$. There is no strict rule for determining the particles that are considered to be the {\it nearest} neighbors. 
Most studies use the concept of the cutoff radius $R_c$ (however, different studies use different values), and some use the Voronoi cell.
We prefer to use the concept of the cutoff radius with $R_c=1.3$, as discussed in section \ref{subsec:rdf}. 
The functions $Y_{lm}(\vb{r}_{ij})$ are the spherical harmonics, $\vb{r}_{ij}$ is the vector connecting the particles $i$ and $j$; $m$ is an integer that runs from $m=-l$ to $m=l$.
Along with $q_l$, $w_l$ parameters are also often used, which can be computed according to formulas:

\begin{equation}
    w_{l}(i) = \frac{ \sum\limits_{m_1+m_2+m_3=0}
    \begin{pmatrix}
    l&l&l \\m_1&m_2&m_3 \end{pmatrix}
    q_{lm_1}(i)q_{lm_2}(i)q_{lm_3}(i)
     }{ \left( \sum\limits_{m=-l}^l \abs{q_{lm}(i)}^2 \right)^{3/2}  } ,
\end{equation}

\noindent where the summation is carried out by the integers $m_1$, $m_2$, $m_3$ from $-l$ to $l$, which satisfy the condition $m_1+m_2+m_3=0$. 
The expression in parentheses is the Wigner 3-j symbol. 
Since no significant conclusions can be drawn from the calculation of the parameters $w_i$, the results for these parameters are presented in supplementary material (subsection \ref{subsubsec:steinhardt_supl}).
To analyze the structure, $q_4$, $q_6$, $q_8$ are most often used. 
In this section, we will present the results for the parameters of test structures with noise, and also apply this method to analyze the structures obtained during our simulation.

\begin{figure}[h]
\center{
    \begin{minipage}[h]{0.47\linewidth}
\center{\includegraphics[width=1\linewidth]{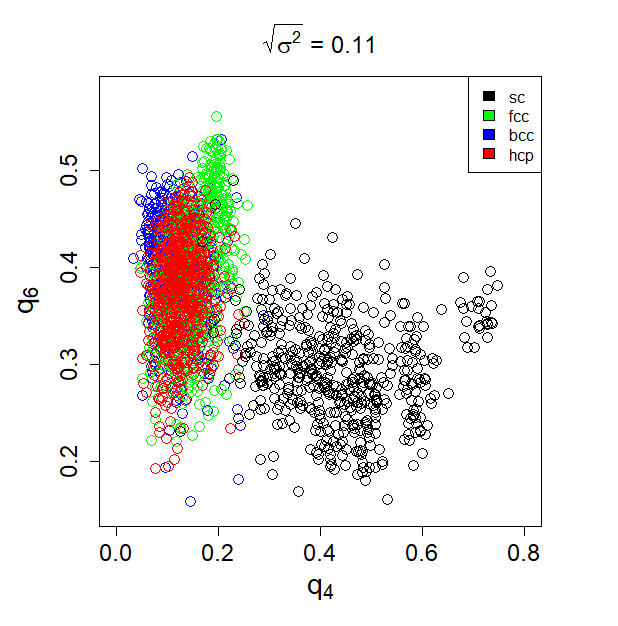} \\ (a)}
\end{minipage}
\begin{minipage}[h]{0.47\linewidth}
\center{\includegraphics[width=1\linewidth]{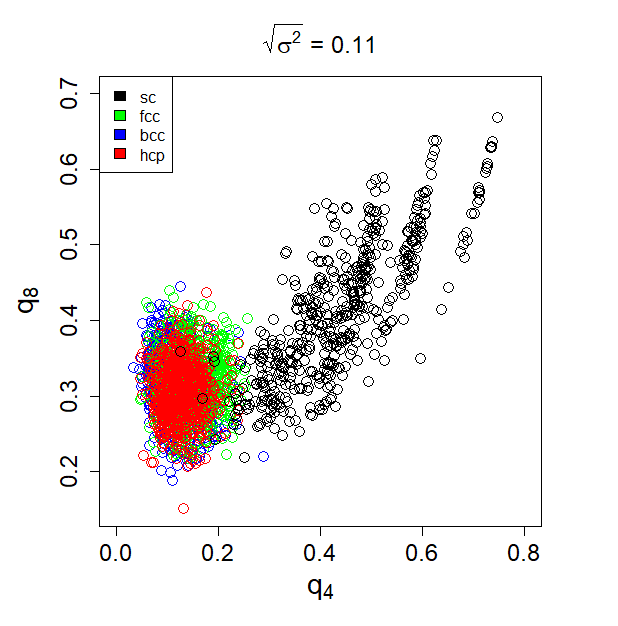} \\ (b)}
\end{minipage}

    \caption{local bond order parameters $q_4-q_6$, $q_4-q_8$ for test structures (a), (b) respectively}
    \label{fig:bop_spots}
    }
\end{figure}

The study of Steinhardt parameters for test structures with the value $\sqrt{\sigma^2}=0.11$ showed that the structures $hcp$, $fcc$, $bcc$ become absolutely indistinguishable (Fig. \ref{fig:bop_spots}). 
This fact does not allow us to apply parameters for our system. 

\begin{figure}[h]
\center{
    \begin{minipage}[h]{0.47\linewidth}
\center{\includegraphics[width=1\linewidth]{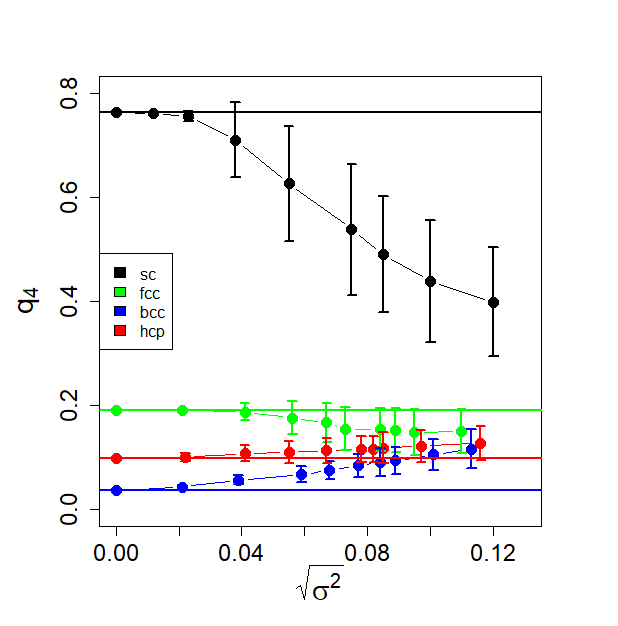} \\ (a)}
\end{minipage}
\begin{minipage}[h]{0.47\linewidth}
\center{\includegraphics[width=1\linewidth]{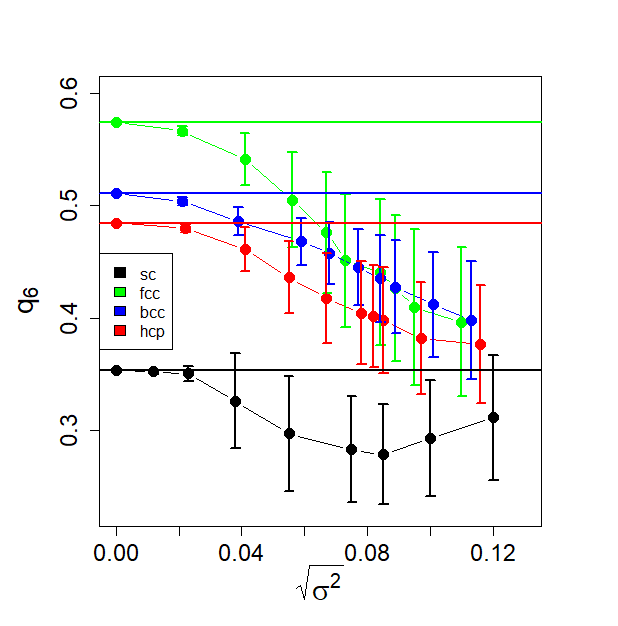} \\ (b)}
\end{minipage}
    \vfill
    \begin{minipage}[h]{0.47\linewidth}
\center{\includegraphics[width=1\linewidth]{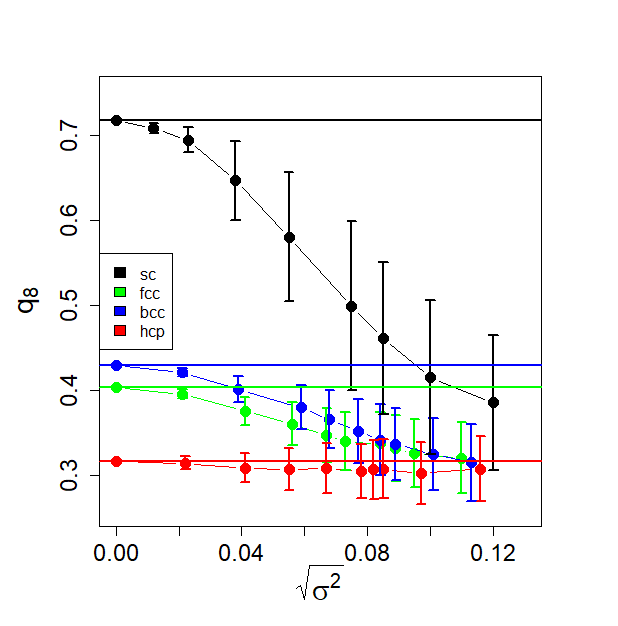} \\ (c)}
\end{minipage}
\begin{minipage}[h]{0.47\linewidth}
\center{\includegraphics[width=1\linewidth]{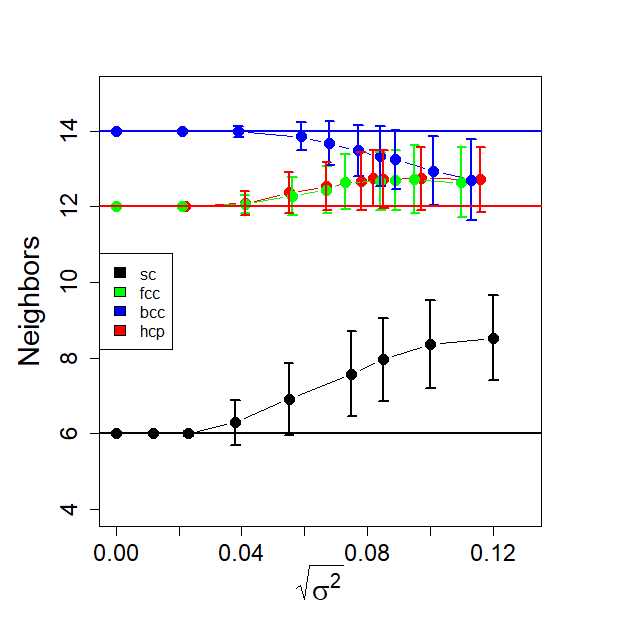} \\ (d)}
\end{minipage}
    \caption{mean values of local bond order parameters $q_4$ (a), $q_6$ (b), $q_8$ (c), number of nearest neighbors (d) for $bcc$ (blue), $hcp$ (red), $fcc$ (green), $sc$ (black) test structures.}
    \label{fig:bop_test_noise}
    
    }
\end{figure}

The study of the dependence of the average parameters on $\sqrt{\sigma^2}$ showed that the mean values with increasing noise deviate greatly from the values for ideal crystal lattices (horizontal lines in Fig. \ref{fig:bop_test_noise}). 
Starting from $\sqrt{\sigma^2}\approx 0.08$, the average number of nearest neighbors also becomes indistinguishable for $bcc$ and $hcp$, $fcc$ and for none of these values corresponds to the true number of neighbors.

Thus, for this model, the application of classical Steinhardt bond order parameters turns out to be impossible. 
In all non-ideal systems, where there is a deviation of the positions of particles (for example, due to thermal fluctuations) from perfect crystal lattices, difficulties arise with determining the crystal structure in simulation. Nevertheless, the potential of the Steinhardt approach is great. 
Spherical invariants are reasonable functions for describing the symmetry of particle clusters.
A number of works have been devoted to modifications of the local bond order parameters to improve the accuracy of determining the type of structures in the presence of thermal fluctuations. 
In this chapter, we will look at several well-known modifications of this method.

\subsubsection{W. Lechner and C. Dellago parameters}
\label{subsubsec:dellago}

In 2008, W. Lechner and C. Dellago \cite{dellago1} proposed a procedure for averaging Steinhardt parameters. 
The authors tested their results on two different systems of soft spheres.  
In this study , they propose to average local bond parameters as follows:

\begin{equation}
    \bar{q}_l(i)=\sqrt{\frac{4\pi}{2l+1}\sum^l_{m=-l} {|\bar{q}_{lm}(i)|}^2 },
\end{equation}

\noindent where 

\begin{equation}
    \bar{q}_{lm}(i) = \frac{1}{\tilde{N}_b(i)} \sum^{\tilde{N}_b(i)}_{k=0} q_{lm}(k). 
\end{equation}

The equation involves a sum over all neighbors of particle $i$, encompassing the particle itself, ranging from $k=0$ to $\tilde{N}_b(i)$. 
To determine the local orientational order vectors for particle $i$, one takes the average of $q_{lm}$ across both the particle $i$ itself and its nearby environment $\tilde{N}_b(i)$. 
While $q_l(i)$ reveals the structural intricacies of the first shell around particle $i$, its averaged counterpart $\bar{q}_l(i)$ accounts for the influence of the second shell. 
The effective considering of the second particle shell is critical in this context.

In the supplementary material (subsection \ref{subsubsec:dellago_supl}) you can also find the results for the updated parameters $\bar{w_l}$ calculated using $\bar{q}_{lm}$:

\begin{equation}
    \bar{w}_{l}(i) = \frac{ \sum\limits_{m_1+m_2+m_3=0}
    \begin{pmatrix}
    l&l&l \\m_1&m_2&m_3 \end{pmatrix}
    \bar{q}_{lm_1}(i)\bar{q}_{lm_2}(i)\bar{q}_{lm_3}(i)
     }{ \left( \sum\limits_{m=-l}^l \abs{\bar{q}_{lm}(i)}^2 \right)^{3/2}  } ,
\end{equation}

Authors \cite{dellago1} conducted their tests on 2 types of soft spheres: with the Lennard-Jones potential and the Gaussian core model. 
We have calculated these parameters for our test structures consisting of solid spheres.

\begin{figure}[h]
\center{
    \begin{minipage}[h]{0.3\linewidth}
\center{\includegraphics[width=1\linewidth]{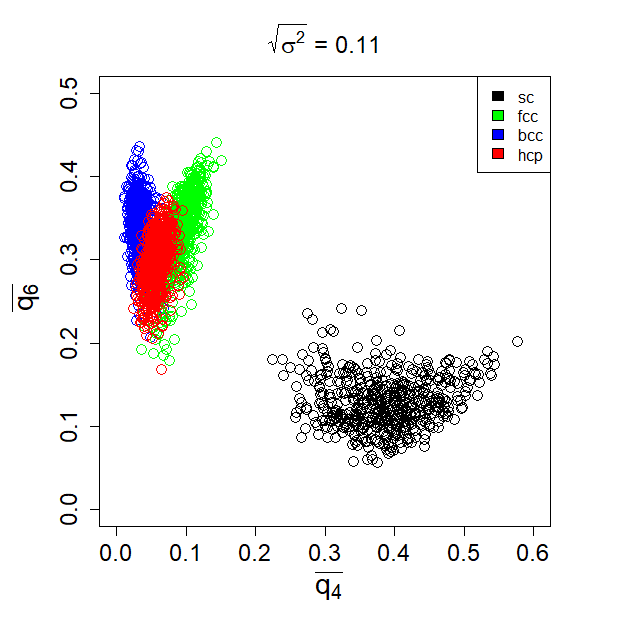} \\ (a)}
\end{minipage}
\begin{minipage}[h]{0.3\linewidth}
\center{\includegraphics[width=1\linewidth]{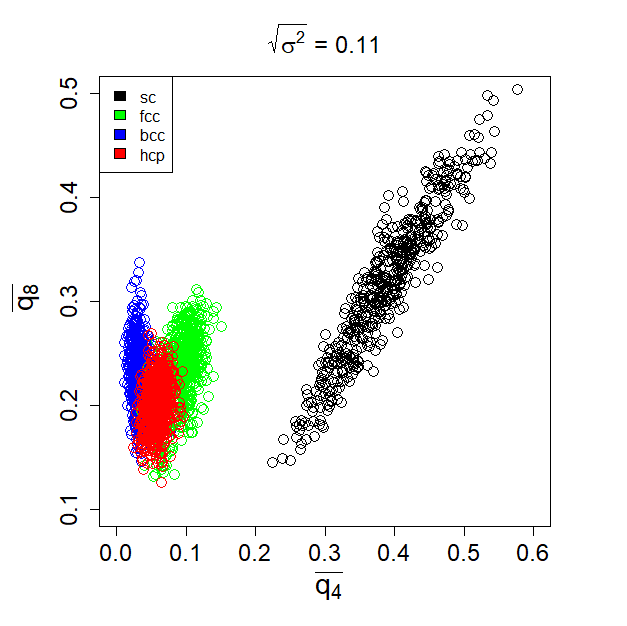} \\ (b)}
\end{minipage}
\begin{minipage}[h]{0.3\linewidth}
\center{\includegraphics[width=1\linewidth]{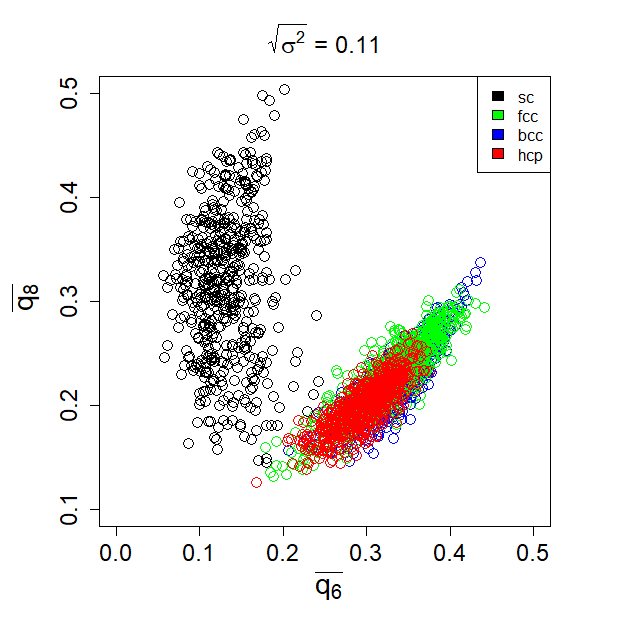} \\ (c)}
\end{minipage}
    \vfill
    \begin{minipage}[h]{0.3\linewidth}
\center{\includegraphics[width=1\linewidth]{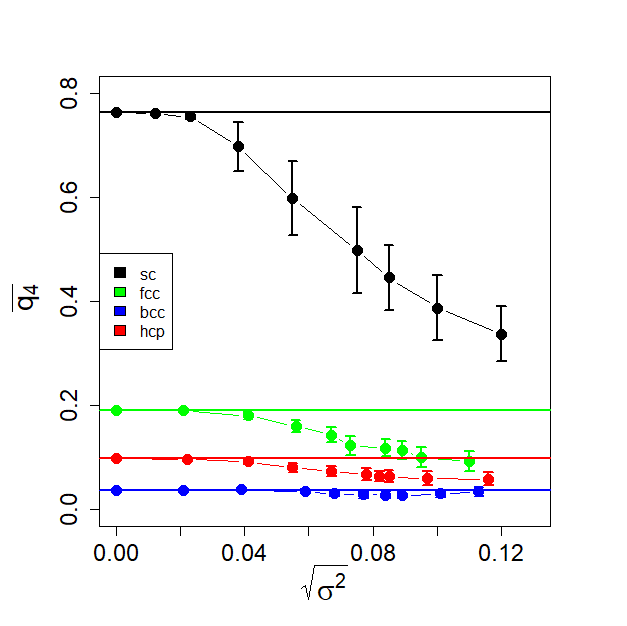} \\ (d)}
\end{minipage}
    \begin{minipage}[h]{0.3\linewidth}
\center{\includegraphics[width=1\linewidth]{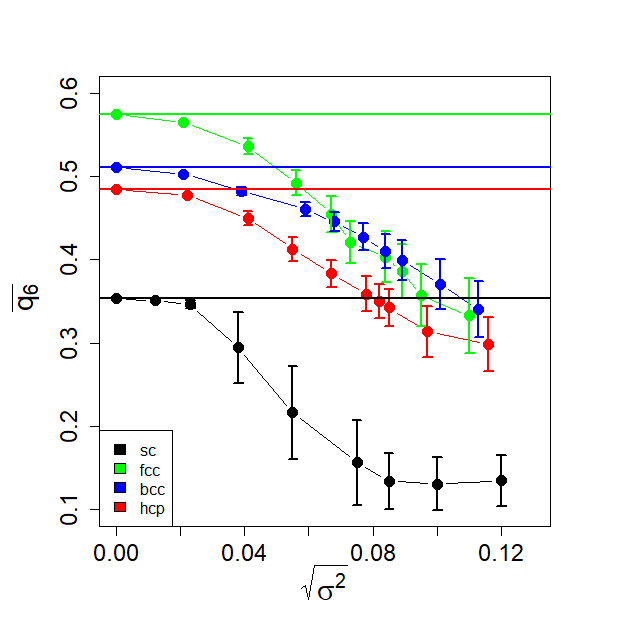} \\ (e)}
\end{minipage}
    \begin{minipage}[h]{0.3\linewidth}
\center{\includegraphics[width=1\linewidth]{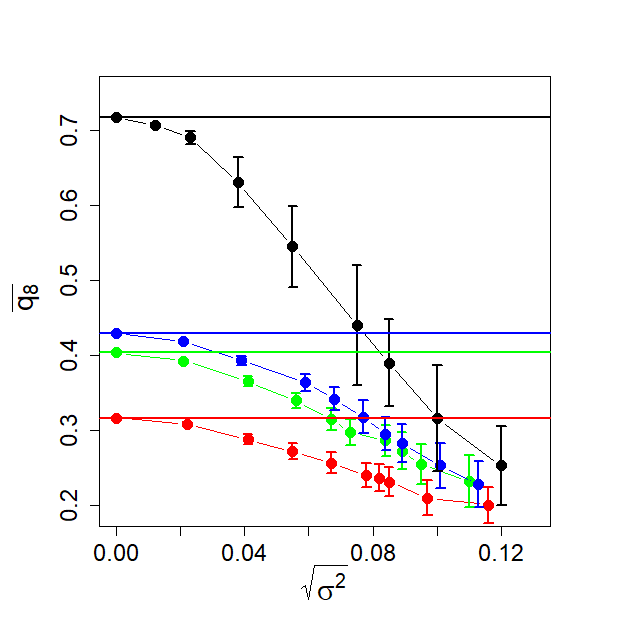} \\ (f)}
\end{minipage}
    \caption{average local bond order parameters $\bar{q}_4-\bar{q}_6$ (a), $\bar{q}_4-\bar{q}_8$ (b), $\bar{q}_6-\bar{q}_8$ (c); mean values of average local bond order parameters $\bar{q}_4$ (d), $\bar{q}_6$ (e), $\bar{q}_8$ (f) for $bcc$ (blue), $hcp$ (red), $fcc$ (green), $sc$ (black) test structures.}
    \label{fig:del_test_noise}
    }
\end{figure}

From the results obtained (Fig. \ref{fig:del_test_noise}) it can be seen that the average values still vary greatly with the growth of $\sqrt{\sigma^2}$, however, averaging definitely reduces the spread in values. 
On the Fig. \ref{fig:del_test_noise} (d) it can be seen that the average values of the parameter $q_4$ are able to distinguish $bcc$, $fcc$, $hcp$, although it is still unreliable. 
The average value for the $bcc$ structure has stabilized and slightly deviates from the undisturbed state, which was not observed with the original method of calculating the Steinhardt parameters Fig. \ref{fig:bop_test_noise} (a). 
The parameters $q_6$, $q_8$ still do not allow us to distinguish structures with noise values comparable to those observed in the system under study ($\sqrt{\sigma}^2\approx 0.11$).

\subsubsection{H. Eslami, P. Sedaghat and F. Müller-Plathe parameters}
\label{subsubsec:mueller}

A more advanced method was proposed by H. Eslami et al. \cite{eslami}. 
In this paper, using the example of the Lennard-Jones system (12-6), a comparison was made with the parameters of previous authors \cite{dellago1}. 
These order parameters examine the ratio of the orientational orders of the second-shell to the first-shell neighbors of a central particle, so the parameters change from 0 (disordered structure) to 1 (in crystal). 
The authors note that despite the fact that the parameters in crystal structures are close to 1, thermal fluctuations reduce the parameters. 
Moreover, it is stronger in $bcc$ than, for example, in $fcc$, due to the large amount of free space and lower density. 
At the same time, no studies have been conducted on the dependence of parameter changes on temperature fluctuations.

These parameters can be calculated using the following formulas:

\begin{equation}
    \tilde{q}_l(i)=\vb{q}_l(i)\cdot\vb{q}_l(j)=\frac{1}{N_b(i)}\sum_{j\in N_b(i)}\sum^l_{m=-l} \hat{q}_{lm}(i)\hat{q}_{lm}^*(j),
\end{equation}

\noindent where

\begin{equation}
    \hat{q}_{lm}(i)=\frac{q_{lm}(i)}{ \Bigl( \sum\limits^l_{m=-l} | q_{lm}(i)|^2 \Bigr)^{1/2} }.
\end{equation}

\noindent After the averaging $\tilde{q}_l(i)$ over the first coordination shell neighbors of particle $i$ one get the following:

\begin{equation}
    \bar{\tilde{q}}_l(i)=\frac{1}{\tilde{N}_b(i)} \sum_{j\in \tilde{N}_b(i)}   \tilde{q}_l(j). 
\end{equation}

Despite the fact that these parameters are not intended to distinguish ideal (noiseless) crystal structures, since they are equal to 1, it is interesting to investigate the dependence of these parameters on noise for systems of solid spheres.
There is a possibility that using a combination of parameters $q_4$, $q_6$, $q_8$ will allow distinguishing different types of structures by the degree of deviation of the parameters from 1. 

\begin{figure}[h]
\center{
    \begin{minipage}[h]{0.3\linewidth}
\center{\includegraphics[width=1\linewidth]{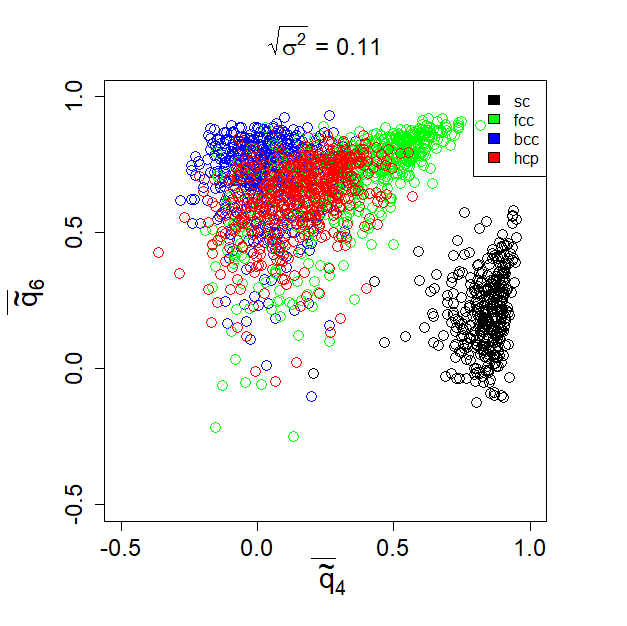} \\ (a)}
\end{minipage}
\begin{minipage}[h]{0.3\linewidth}
\center{\includegraphics[width=1\linewidth]{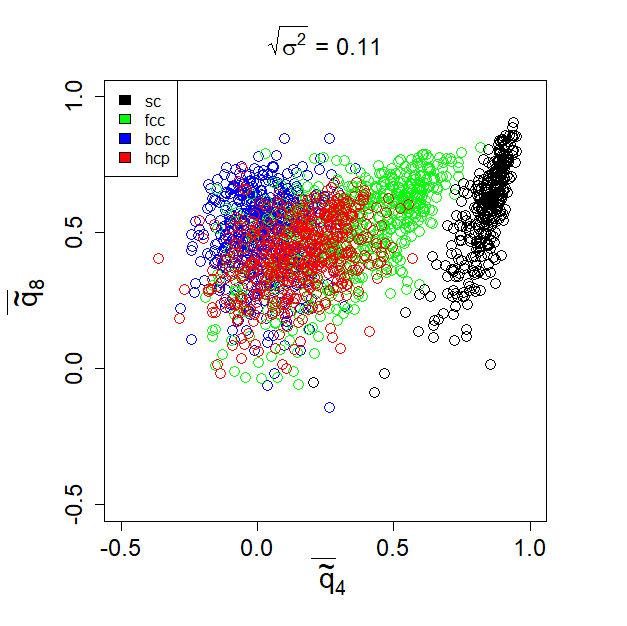} \\ (b)}
\end{minipage}
\begin{minipage}[h]{0.3\linewidth}
\center{\includegraphics[width=1\linewidth]{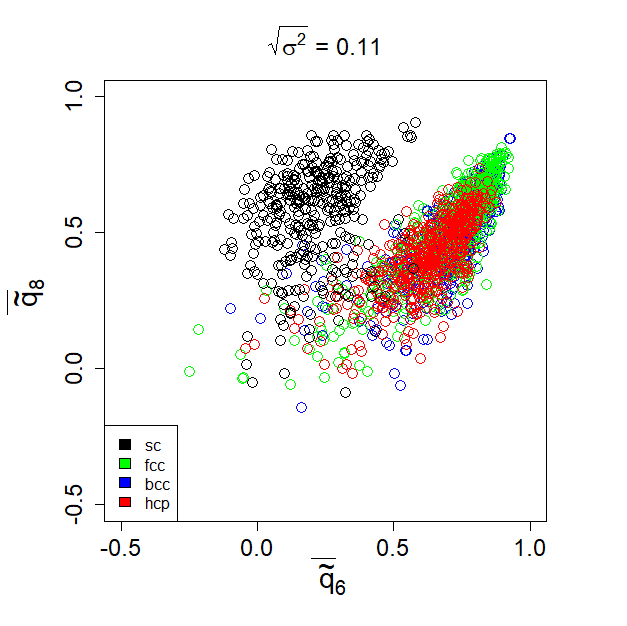} \\ (c)}
\end{minipage}
    \vfill
    \begin{minipage}[h]{0.3\linewidth}
\center{\includegraphics[width=1\linewidth]{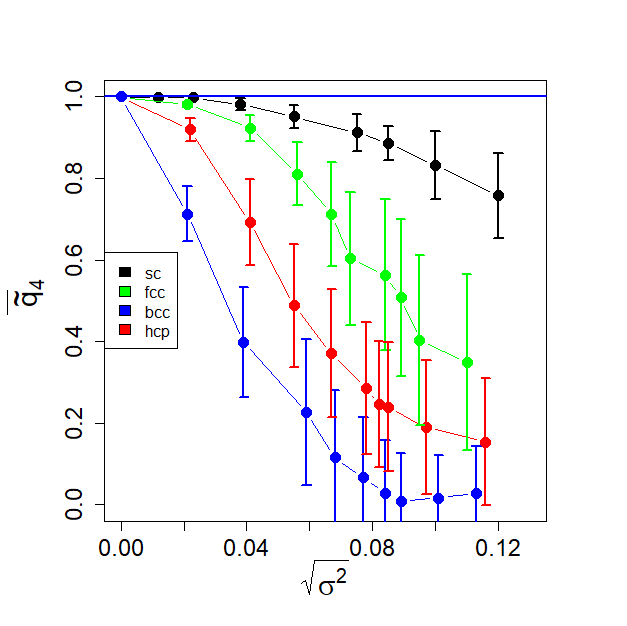} \\ (d)}
\end{minipage}
    \begin{minipage}[h]{0.3\linewidth}
\center{\includegraphics[width=1\linewidth]{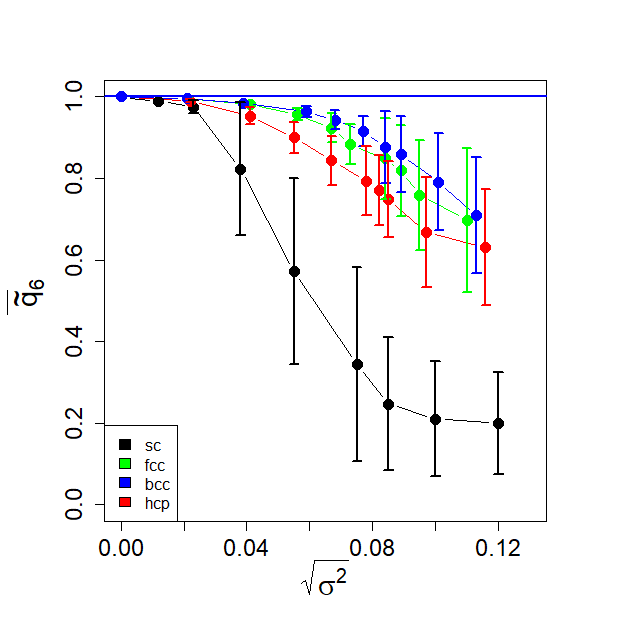} \\ (e)}
\end{minipage}
    \begin{minipage}[h]{0.3\linewidth}
\center{\includegraphics[width=1\linewidth]{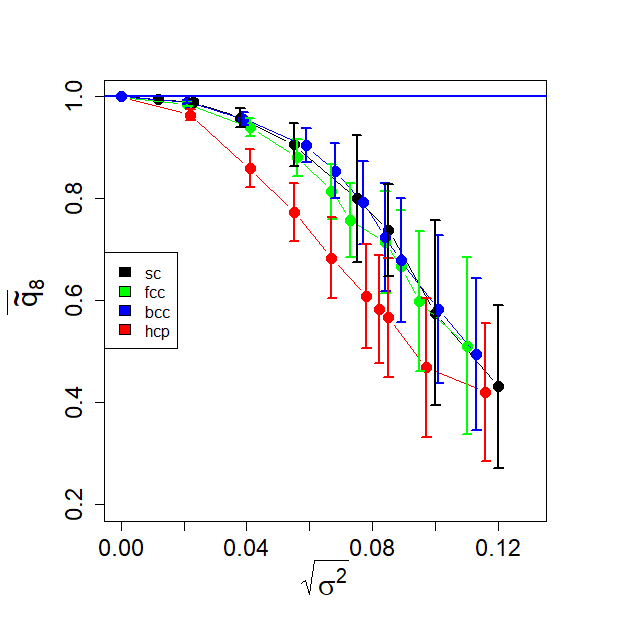} \\ (f)}
\end{minipage}
    \caption{local bond order parameters $\bar{\tilde{q}}_4-\bar{\tilde{q}}_6$ (a), $\bar{\tilde{q}}_4-\bar{\tilde{q}}_8$ (b), $\bar{\tilde{q}}_6-\bar{\tilde{q}}_8$ (c); mean values of local bond order parameters $\bar{\tilde{q}}_4$ (d), $\bar{\tilde{q}}_6$ (e), $\bar{\tilde{q}}_8$ (f) for bcc (blue), hcp (red), fcc (green), sc (black) test structures.}
    \label{fig:mueller_test_noise}
    }
\end{figure}

From the obtained data (Fig. \ref{fig:mueller_test_noise}), it can be seen that the parameter $\bar{\tilde{q}}_4$ still seems to be the most sensitive for separating structures. 
Depending on the type of structure, noise has a different effect on the deviation of parameters from undisturbed values.
Such a wide range of values, unfortunately, does not allow using combinations of parameters $q_4$, $q_6$, $q_8$ to distinguish between different types of structures.

Thus, since the parameters discussed in the section \ref{subsec:local} do not allow us to reliably distinguish test structures with a noise value comparable to that estimated in the system under study, we cannot rely on them when analyzing the system under study. 
In the next section, another way of using Steinhardt parameters (Eq. \ref{stein_eq}) will be proposed, the efficiency of the method on test structures will be checked, and the use of the method on the example of the system under study will also be demonstrated.
  \subsection{Noise reduction procedure}
\label{subsec:noise_reduction}

As we have established in the previous sections, the methods used for analyzing structures do not allow us to distinguish satisfactorily different symmetries in the presence of noise. 
Therefore, at this stage of the work, an attempt was made to reduce the influence of noise by averaging the position of coordinates in space.

\begin{figure}[h]
\center{
    \begin{minipage}[h]{0.4\linewidth}
\center{\includegraphics[width=1\linewidth]{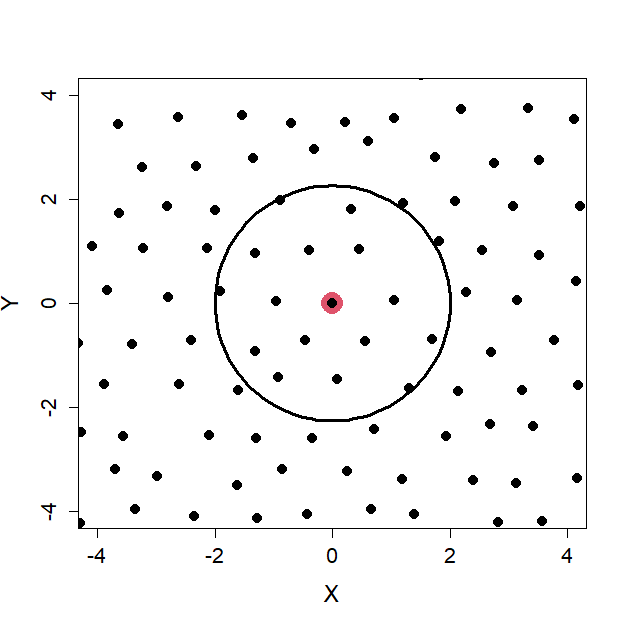} \\ (a)}
\end{minipage}
\begin{minipage}[h]{0.4\linewidth}
\center{\includegraphics[width=1\linewidth]{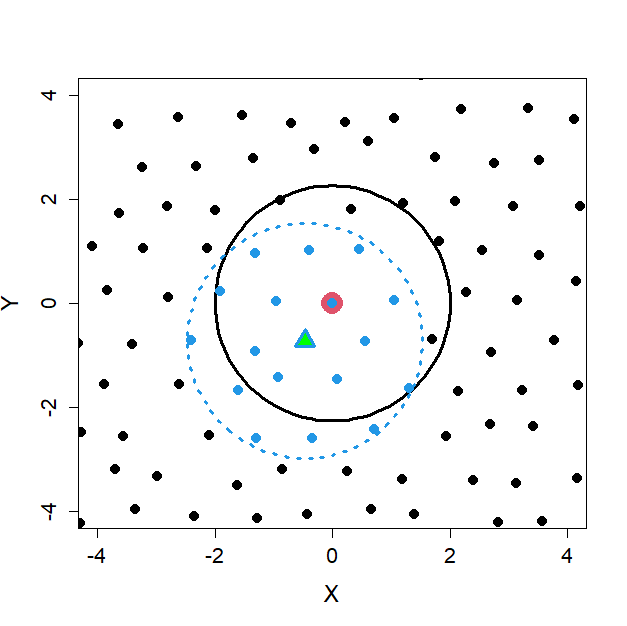} \\ (b)}
\end{minipage}
    \vfill
    \begin{minipage}[h]{0.4\linewidth}
\center{\includegraphics[width=1\linewidth]{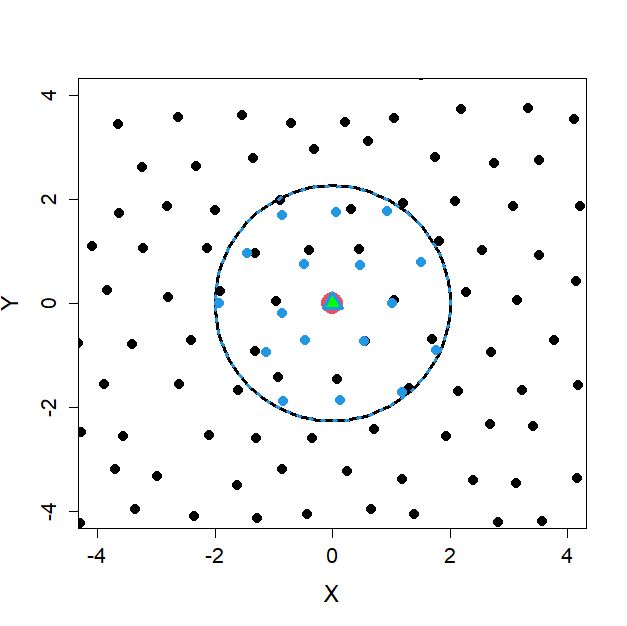} \\ (c)}
\end{minipage}
\begin{minipage}[h]{0.4\linewidth}
\center{\includegraphics[width=1\linewidth]{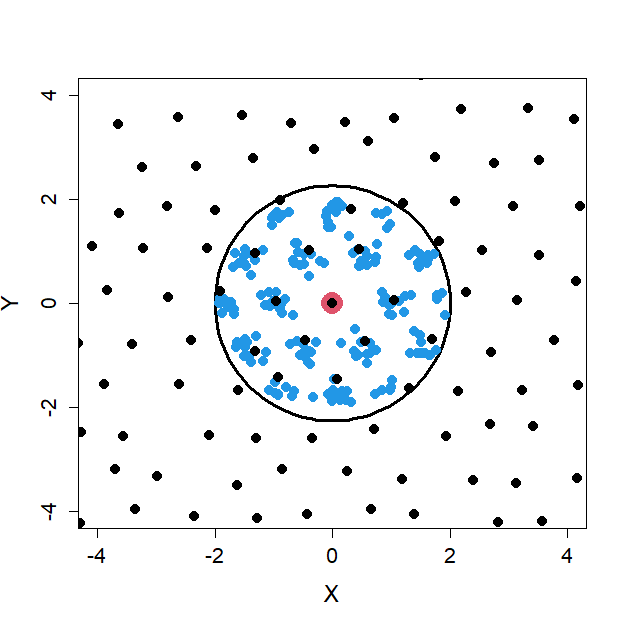} \\ (d)}
\end{minipage}
    \caption{two-dimensional coordinate averaging scheme. (a) Analyzed red particle $i$ with its neighbors in black sphere; (b) triangle particle $j$ from the original sphere and its blue neighbors; (c) blue sphere is completely translated into the original black one; (d) steps (b)-(c) are done for all neighbors from the black sphere.
    \label{fig:av_proc}
    }
}  
\end{figure}

We propose the following procedure (Fig. \ref{fig:av_proc}). 
Let's choose the particle $i$ for which one will compute Steinhardt order parameters. 
We draw a sphere of radius $R_s=2$ around the particle (Fig. \ref{fig:av_proc} a). 
Let it have $N_i$ neighbors of particle $i$. Next, select particle $j$ (green triangle on Fig. \ref{fig:av_proc} b) from the list of neighbors $\{N_i\}$. 
We also draw a sphere of radius $R_s=2$ around it, into which $N_j$ particles fall (blue particles on Fig. \ref{fig:av_proc}b). 
Then we translate particle $j$ together with all its neighbors into particle $i$ (Fig. \ref{fig:av_proc} c). 
We perform the same operation with all the particles from the list $\{N_i\}$ (Fig. \ref{fig:av_proc} d). 

The next step after moving the points inside the sphere is averaging their positions. 
To do this, we set the positions of the neighbors of the initial particle as the starting position of the groups $\vb{r'}_{gr}$ (black dots inside the black sphere on the Fig.\ref{fig:av_proc} a)). 
Then we iteratively refine the present position of the centers of the groups $\vb{r}_{gr}$.  To do this, add the position of each of the above points (blue on the Fig. \ref{fig:av_proc} d)) to each of the existing groups as an gaussian term. To get the updated coordinates of the groups, it is necessary to weigh the amounts received.

\begin{equation}
    \vb{r}_{gr} = \frac{1}{Z_{gr}}\sum_k \vb{r}_k  \exp{-\frac{\left(\vb{r}_k-\vb{r'}_{gr}\right)^2}{2\cdot 2\sigma^2}}
\end{equation}
\begin{equation}
    Z_{gr} = \sum_k  \exp{-\frac{\left(\vb{r}_k-\vb{r'}_{gr}\right)^2}{2\cdot 2\sigma^2}},
\end{equation}

\noindent where $\vb{r}_k$ is the position of the particle relative to the particle under consideration $i$, $\vb{r'}_{gr}$ are group positions at previous iteration step, $\vb{r}_{gr}$ are updated group positions, $2\sigma^2 = 0.0242$, as defined in the section \ref{subsec:noise_estimation}.
We repeat this procedure a number of times, achieving convergence of the position of the groups. 
Thus, we obtained the averaged positions of the neighbors of the initial point $i$, which we use to calculate the Steinhardt parameters (Eq. \ref{stein_eq}). 

It should be noted that for averaging coordinates, the radius of the sphere ($R_s=2$) is set obviously larger than the cutoff radius for calculating the Steinhardt parameters, which is still equal to $r_c=1.3$. 
During iterations, the coordinates of the centers of the groups $\vb{r}_{gr}$, which will be used as the coordinates of neighbors at the end, may shift. 
Thus, if the original point $i$ had an insufficient or excessive number of points inside the sphere $R_c=1.3$, in the process of iterations, this amount changes and becomes less sensitive to noise around the original particle. Thus, the choice of $R_s=$2 provides a more reliable averaging.

\begin{figure}[ht]
\center{
    \begin{minipage}[h]{0.3\linewidth}
\center{\includegraphics[width=1\linewidth]{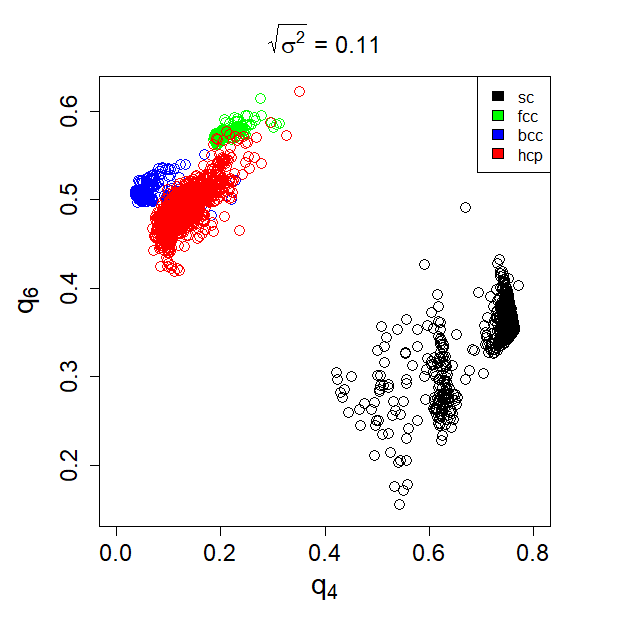} \\ (a)}
\end{minipage}
\begin{minipage}[h]{0.3\linewidth}
\center{\includegraphics[width=1\linewidth]{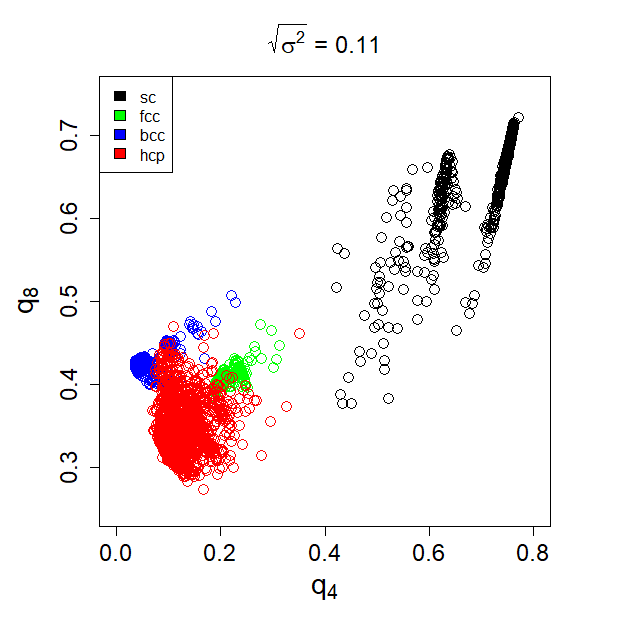} \\ (b)}
\end{minipage}
\begin{minipage}[h]{0.3\linewidth}
\center{\includegraphics[width=1\linewidth]{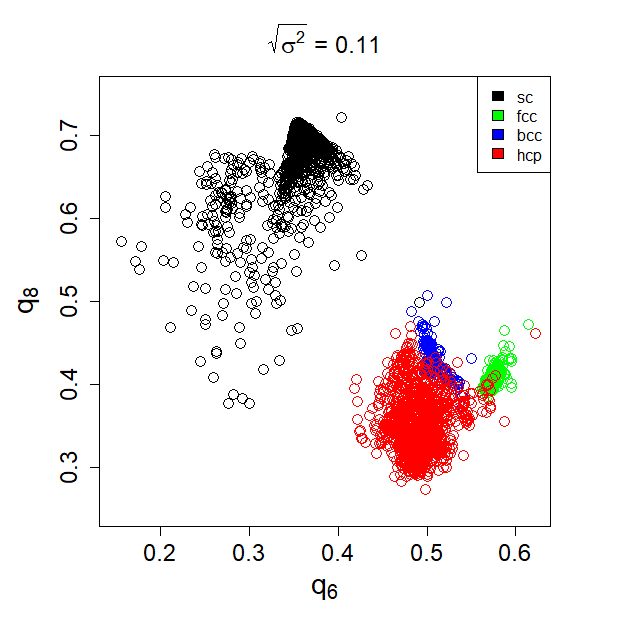} \\ (c)}
\end{minipage}
    \vfill
    \begin{minipage}[h]{0.3\linewidth}
\center{\includegraphics[width=1\linewidth]{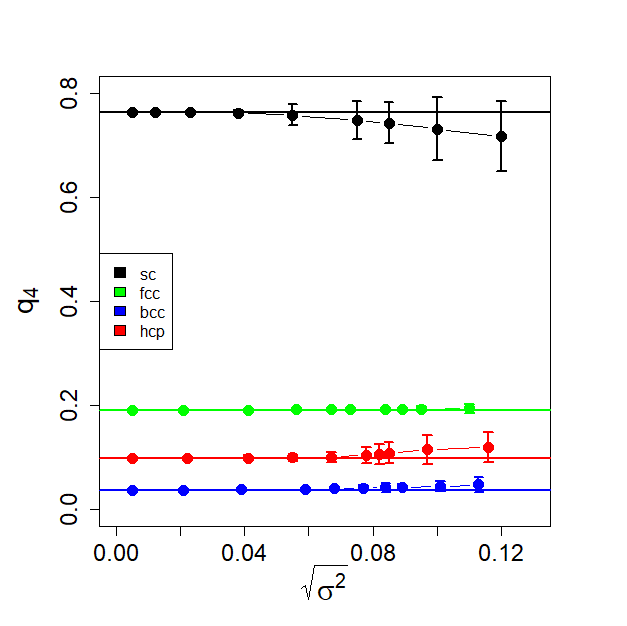} \\ (d)}
\end{minipage}
    \begin{minipage}[h]{0.3\linewidth}
\center{\includegraphics[width=1\linewidth]{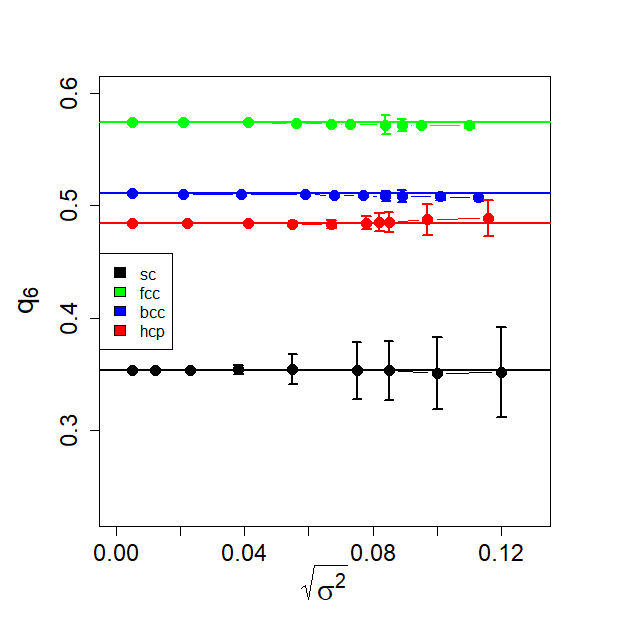} \\ (e)}
\end{minipage}
    \begin{minipage}[h]{0.3\linewidth}
\center{\includegraphics[width=1\linewidth]{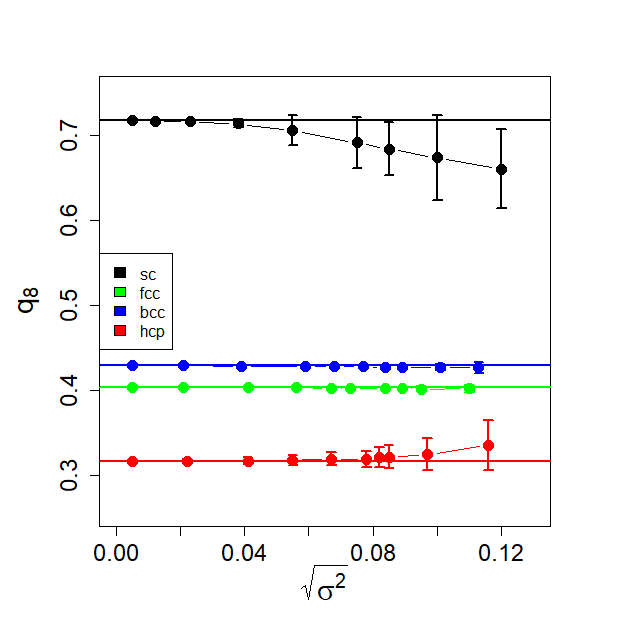} \\ (f)}
\end{minipage}
    \caption{local bond order parameters after averaging procedure $q_4-q_6$ (a), $q_4-q_8$ (b), $q_6-q_8$ (c); mean values of local bond order parameters $q_4$ (d), $q_6$ (e), $q_8$ (f) for $bcc$ (blue), $hcp$ (red), $fcc$ (green), $s$c (black) test structures.}
    \label{fig:test_reduce_noise}
    }
\end{figure}

As we can see from the results obtained (Fig. \ref{fig:test_reduce_noise}), the procedure of averaging coordinates in space makes it possible to make the mean values of all parameters more stable with increasing noise. 
At the same time, the spread of order parameters naturally decreases. 
On the planes $q_4$-$q_6$, $q_4$-$q_8$, $q_6$-$q_8$ (Fig. \ref{fig:test_reduce_noise} a - c) it is also clearly seen that when introducing displacements in the lattice coordinates of the order $\sqrt{\sigma^2} = 0. 11$ the structures are clearly distinguishable. 
The same pictures show the division of $sc$ into 3 groups. 
It is established that the reason for this is the different number of neighbors. 
Thus, one can conclude that the combination of three parameters $q_4$, $q_6$, $q_8$ allows to distinguish different symmetries after the averaging procedure, even if there is a sufficiently large noise comparable to that observed in the system under study.

Another modification of this method was also tested. Instead of combining the particles of the black sphere during averaging procedure by translations (Fig.\ref{fig:test_reduce_noise} a), the coordinates of the centers of mass for each of the spheres (black and all blue ones) were superimposed. 
Thus, one can say that the displacement of the central red particle is also taken into account. 
However, our experience has shown that this modification does not provide significant improvements. 
Initially, the described method can be considered optimal, and it will be used in further analysis. 
The results for the superposition of the centers of mass can be found in supplementary materials (\ref{subsec:noise_reduction_supl}).

\begin{figure}[h]
\center{
    \begin{minipage}[h]{0.45\linewidth}
\center{\includegraphics[width=1\linewidth]{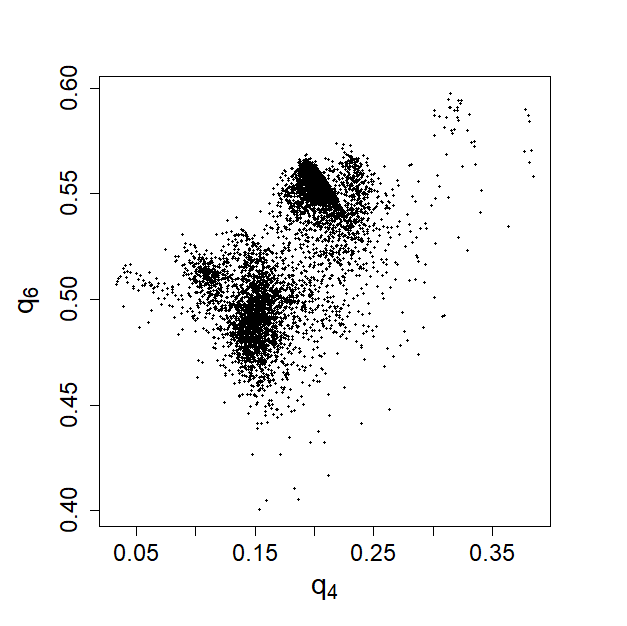} \\ (a)}
\end{minipage}
\begin{minipage}[h]{0.45\linewidth}
\center{\includegraphics[width=1\linewidth]{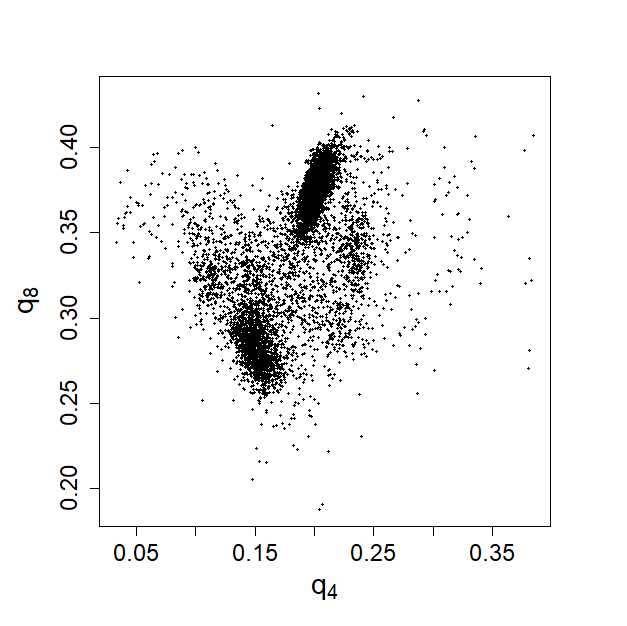} \\ (b)}
\end{minipage}
    \vfill
    \begin{minipage}[h]{0.45\linewidth}
\center{\includegraphics[width=1\linewidth]{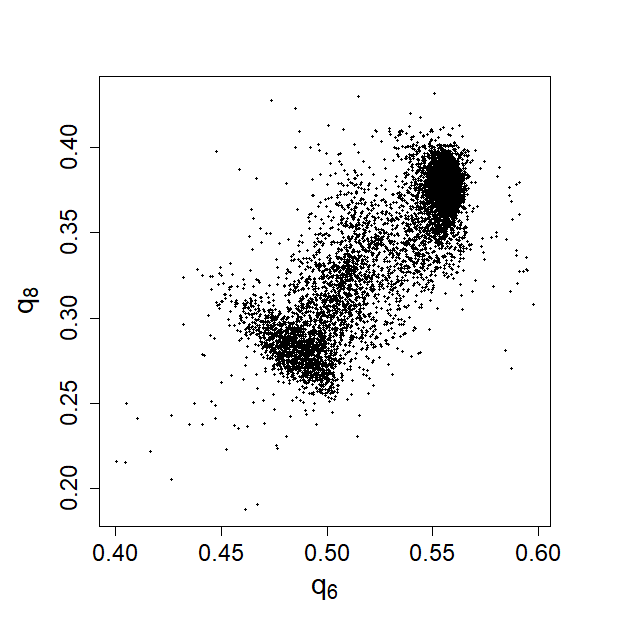} \\ (c)}
\end{minipage}
\begin{minipage}[h]{0.45\linewidth}
\center{\includegraphics[width=1\linewidth]{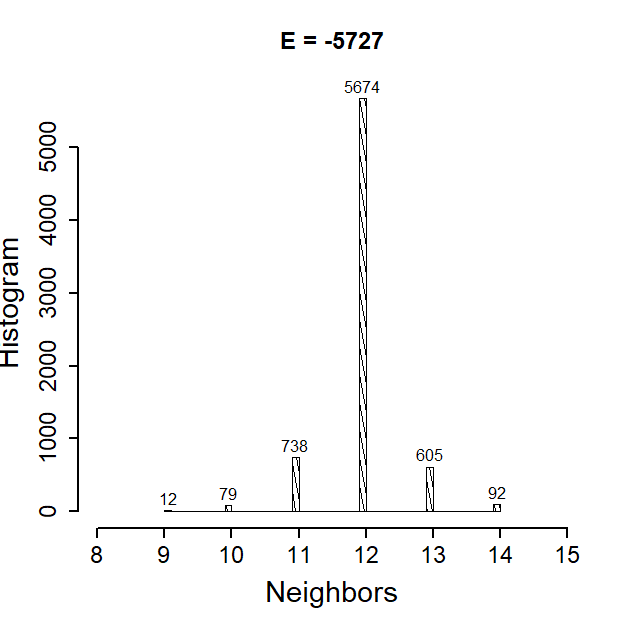} \\ (d)}
\end{minipage}
    \caption{local bond order parameters for the bulk system $E=-5727$ after averaging coordinates: $q_4$-$q_6$ (a),$q_4$-$q_8$ (b),$q_6$-$q_8$ (c), neighbors histogram (d)
    \label{fig:separ_step0}
    }
}  
\end{figure}

Since the proposed approach has proven itself well in relation to test structures, we have every reason to apply it to the system under study. 
To begin with, let's take a system without walls with low energy $E=-5727$ (Fig. \ref{fig:bulk5727}). 
The obtained parameters after the averaging procedure are shown in Fig. \ref{fig:separ_step0}.

From the data obtained, it can be seen that the parameters $q_4$, $q_6$, $q_8$ are clearly divided into several groups. 
Two dense clusters are allocated on the plane $q_6$-$q_8$. 
On the other two planes, smaller clumps are also noticeable. 
However, since their size is small and they are located close to larger ones, we will consider them part of larger clusters. 
It was also found that most of the particles in two biggest the most dense clusters have 12 neighbors, while scattered points have a predominantly different number of neighbors than 12. 
Since the predominance of the correct number of neighbors can be observed on the test structures after the averaging procedure, then at this stage we can assert that there is a coexistence of 2 types of symmetries with 12 neighbors in this structure, while a different number of neighbors is formed due to the noise of the structure. 
Here we make the assumption that these 2 crystal structures originate at higher energies, where they coexist with the disordered phase, which we also will call $Melt$. 
We present an algorithm for isolating co-existing structures at different energies.

{\sf{\it Procedure for structure separation}}

\underline{Step 1}

At the first stage, we calculated the parameters for the averaged coordinates of a low-energy system ($E=-5727$) and divided the particles into 2 clusters using the k-means method \cite{kmeans} built into the R environment. 
$q_4$, $q_6$, $q_8$ were used as input data for clustering to increase accuracy. 
At the same time, our experience has shown that additional consideration of neighbors as input data does not significantly affect the result.
The number of clusters (2) and the number of launches ($n_{start} = 500, 1000$) are used as external parameters for the method. 
The main disadvantage of this method is a strong dependence on the starting points chosen randomly. 
To get around this problem, one should do several launches, and it is also better to visually observe the results of the cluster separation on the graphs. 
Despite the aforementioned drawback, within the framework of our task, the results are well reproducible and do not change when restarted. 
One can also use other clustering methods (for example, various variations of the k-means method \cite{kmeans_var}, the c-means method, EM-clustering). 
Each of these methods has its own advantages and disadvantages, however, in our work we will not describe and compare clustering algorithms in detail.
We prefer to use k-means method because of its simplicity, but testing other methods has yielded similar results.
After the initial separation of the parameters of the low-energy structure into 2 clusters, it is necessary to clarify the positions of the clusters centers in the space {$q_4, q_6, q_8$}. 
To take into account the asymmetry of the clusters, we calculate the covariance matrix:

\begin{equation}
    \hat{C}_{k} = \begin{bmatrix}
    cov\:( q_4^{k},\; q_4^{k} ) & cov\:( q_4^{k}, \;q_6^{k} ) & cov\:( q_4^{k}, \;q_8^{k} )  \\
    cov\:( q_6^{k},\; q_4^{k} ) & cov\:( q_6^{k},\; q_6^{k} ) & cov\:\:( q_6^{k},\; q_8^{k} )  \\
    cov\:( q_4^{k},\; q_8^{k} ) & cov\:( q_6^{k}, \;q_8^{k} ) &cov\:( q_8^{k},\; q_8^{k} ) 
    \end{bmatrix},
    \label{covar}
\end{equation}

\noindent here $k = \{1,2\}$ is the number of cluster; $q_j^k$ - the parameter of the particle related to the cluster $k$; $j = \{4,6,8\}$ is the order of the parameter. 
Initially, the mean value of the parameter in the initial cluster is taken as the center of the cluster, then this value will be refined during the iterative process. 
In the iterative process the weight with which each particle $i$ of the structure enters each cluster $k$ is calculated:

\begin{equation}
    \Omega_i^k = \sqrt{det \: \hat{C}_k^{-1}} \exp{ -\frac{1}{2} ({q}_{i,4} -\bar{q}_4^{k}, \: {q}_{i,6} - \bar{q}_6^{k}, \: {q}_{i,8} - \bar{q}_8^{k} ) \;\; \hat{C}_k^{-1}\begin{pmatrix} {q}_{i,4} -\bar{q}_4^{k} \\ {q}_{i,6} -\bar{q}_6^{k} \\ {q}_{i,8} -\bar{q}_8^{k} \end{pmatrix}  },  
    \label{weights}
\end{equation}

\noindent where $\hat{C}_k^{-1}$ is the inverse matrix of $\hat{C}_k$; $\bar{q}_j^k$ - the center of cluster $k$. 
Then the particles are reassigned to clusters: each particle $i$ belongs to the cluster $k$ for which the weight turned out to be greater. 
After that, the positions of cluster centers are updated as:

\begin{equation}
\bar{q}_j^{k} = \frac{ \sum\limits_i q_{i,j}^{k} \; \Omega_i^{k} }{ \sum\limits_i \Omega_i^{k}},   
\end{equation}

\noindent where $j=\{4,6,8\}$; $i$ - the number of particle form the cluster $k$. 
The iterative process ends at the moment when $\bar{q}_j^{k}$ stops to change. 

{\it Note:} 
The $\hat{C}_k$ and $\hat{C}_k^{-1}$ do not change during the iterative process. 
In the case of matrix recalculation, we will get nested clusters with close centers.

Thus, at the Step 1, the centers of 2 crystal structures ($\bar{q}_j^{k}$, $k=\{1,2\}$, $j=\{4,6,8\}$ ) and covariance matrices ($\hat{C}_k$, $k=\{1,2\}$) for them were obtained. 
These values are fixed and do not change in further steps.

\underline{{\it Step 2.}}

After 2 types of particles with different symmetries have been determined for this structure, we propose that these types are formed at higher energies and coexist with the polymer melt. 
For the 2 found crystal states we fix the found values of cluster centers, as well as covariance matrices. 
At this step, our task is to find the values ($\bar{q}_4^3$, $\bar{q}_6^3$, $\bar{q}_8^3$, $\hat{C}_3$) corresponding to the polymer melt. 
To do this, we choose a high-energy structure, such that only a melt is observed. 
For example, let's take the energy $E = -2040$. 
For simplicity, assuming that there are no crystalline particles in the sample, we find $\bar{q}_4^3$, $\bar{q}_6^3$, $\bar{q}_8^3$ in the sample as the mean parameters for the entire system:

\begin{equation}
\bar{q}_j^{3} = \frac{1} {N \cdot N_c} \sum\limits_i q_{i,j} ,   
\end{equation}
\noindent where $N\cdot N_c$ is the number of particles in the system ($N \cdot N_c = 7200$ in small and $N\cdot N_c = 14400$ in large systems, respectively).

And based on these data, we build a covariance matrix (Eq. \ref{covar}).

Thus, at the Step 2, the center of disordered structure ($\bar{q}_j^{3}$, $j=\{4,6,8\}$ ) and covariance matrix ($\hat{C}_3$) for it was obtained. 
These values are fixed and do not change in further step. 
One can find results for $\bar{q}_j^{k}$ ($k = \{1,2,3\}$) in the Tab. \ref{tab:found_param}.

\begin{figure}[h]
\center{
\center{\includegraphics[width=0.55\linewidth]{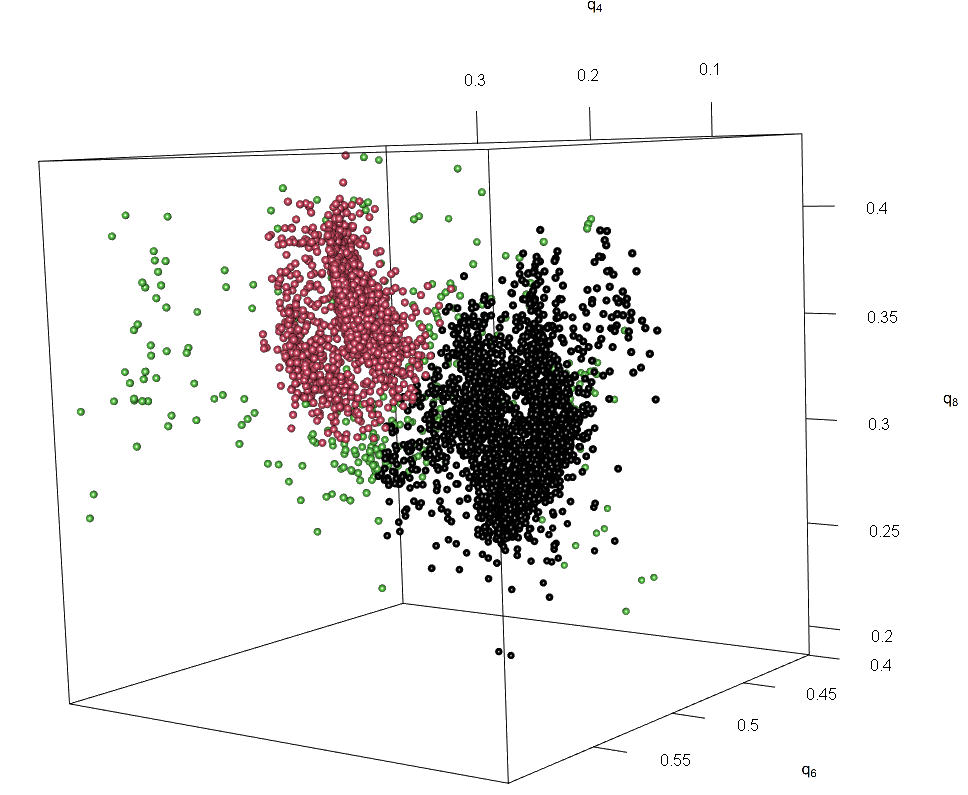} \\ }
    \caption{3d view of Steinhardt bond order parameters for the system without walls at $E=-5727$, $L_x=L_y=20, L_z=19$
    \label{fig:stein3d_bulk5727}
    }
}  
\end{figure}

\underline{{\it Step 3.}}

Since we know the coordinates in the \{$q_4$, $q_6$, $q_8$\} space for the three phases and their covariance matrices, it is now possible to distribute the particles in the system under study of any energy into 3 types. 
First of all, we can return to the low-energy structure and redistribute using weights (Eq. \ref{weights}) the particles over the 3 sets of $\{ \bar{q}_4, \bar{q}_6, \bar{q}_8 \}$ found. 
Since the system we have chosen does not have the lowest of all possible energies ($E = -5760$), the presence of several isotropic phase particles is possible.

\underline{Remark}

To make sure that the sets obtained in the Tab. \ref{tab:found_param} for these 3 clusters do not depend significantly on the choice of specific structures, additional tests were conducted. 
The procedure of searching $\bar{q}_j^{k}$ and $\hat{C}_k$ ($j=\{4,6,8\}$, $k=\{1,2,3\}$) was repeated for several sets of low and high energy conformations. 
The set of conformations obtained during $SAMC$ can be considered independent, since they were taken with a large time difference (on the order of several months of computer calculations). 
The results (Step 1, Step 2) of several tests for the system described here can be found in the supplementary material \ref{subsec:noise_reduction_supl}. The values $\bar{q}_j^{k}$ and $\hat{C}_k$ do not depend significantly on the choice of conformations. 
The assignment of a part to different clusters (Step 3) also does not change significantly.

\begin{figure}[h]
\center{
    \begin{minipage}[h]{0.32\linewidth}
\center{\includegraphics[width=1\linewidth]{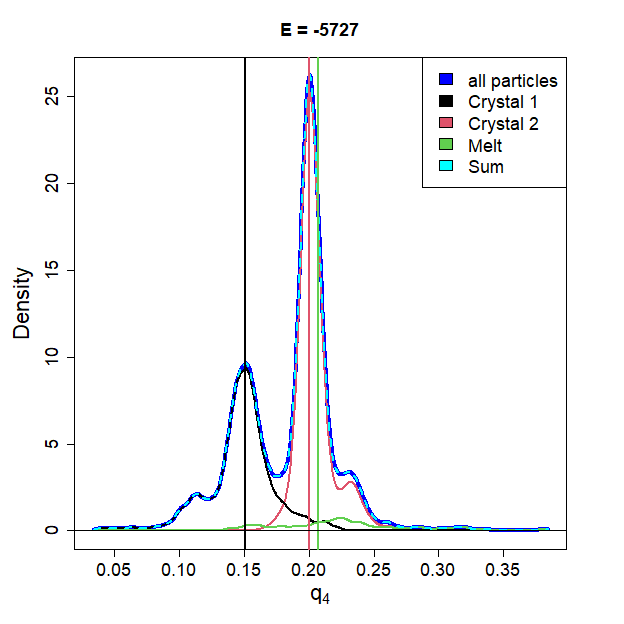} \\ (a)}
\end{minipage}
\begin{minipage}[h]{0.32\linewidth}
\center{\includegraphics[width=1\linewidth]{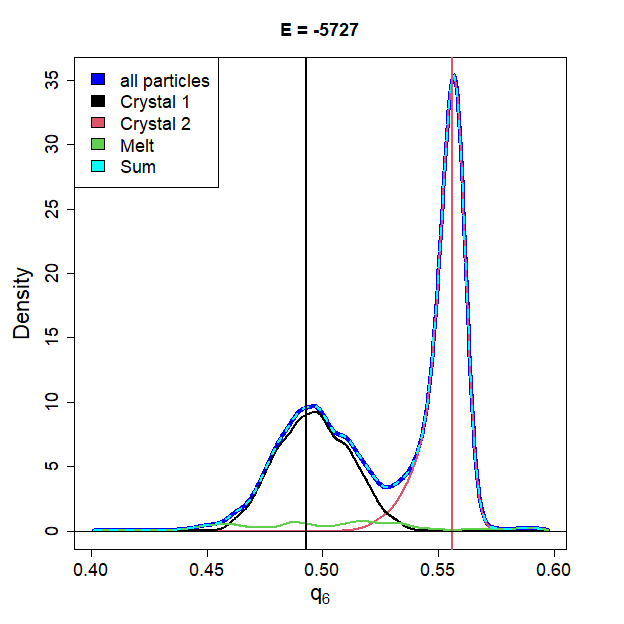} \\ (b)}
\end{minipage}
    \begin{minipage}[h]{0.32\linewidth}
\center{\includegraphics[width=1\linewidth]{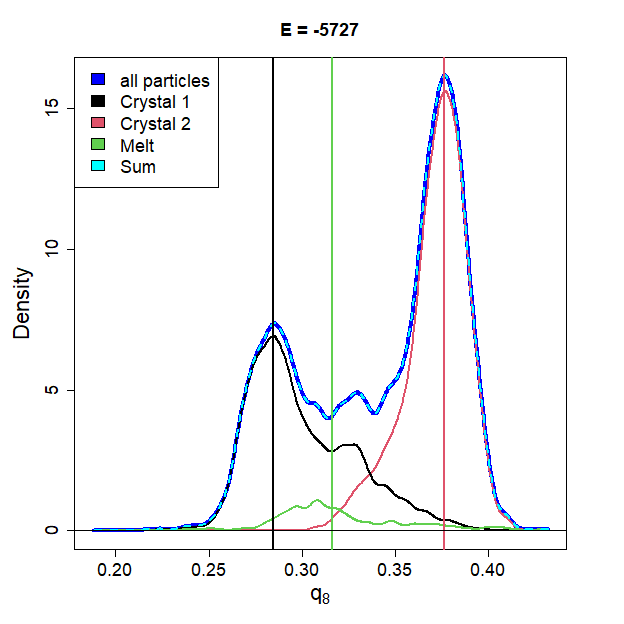} \\ (c)}
\end{minipage}
    \caption{local bond order parameters distribution for the bulk system $E=-5727$ , $L_x=L_y=20$, $L_z=20$ after averaging coordinates
    \label{fig:separ_dens}
    }
}  
\end{figure}

In the Fig. \ref{fig:separ_dens}, the distributions of the parameters $q_4$, $q_6$, $q_8$ for the system under study are shown in blue. 
Black, red and green represent the distributions of parameters weighted by the fraction of particles of the corresponding cluster. 
The dotted line of the cyan color represents the sum of these individual contributions. 
It is obvious that the distribution of parameters for all particles is actually a set of distributions of particles belonging to individual clusters, since the blue curve exactly coincides with the cyane line. 
The shoulder in the black curve at $q_8$ distribution for $Crystal\;1$ (Fig. \ref{fig:separ_dens} c) corresponds to the number of neighbors other than 12 (detailed distributions can be seen in the additional materials \ref{subsec:noise_reduction_supl}).

\begin{table}[ht]
\center{
\begin{tabular}{|c|c||c|c|c|}
\hline
 & color & $\bar{q}_4$ & $\bar{q}_6$ & $\bar{q}_8$\\
\hline 
\hline
Crystal 1 & black & $0.149 \pm 0.030$ & $0.491 \pm 0.017$ & $0.288 \pm 0.028$ \\
\hline
Crystal 2 & red & $0.201 \pm 0.021$ & $0.556 \pm 0.011$ & $0.376 \pm 0.020$ \\
\hline
Melt & green & $0.223 \pm 0.066$ & $0.401 \pm 0.082$ & $0.333 \pm 0.053$\\
\hline

\end{tabular}
    \caption{parameters $\bar{q}_4$, $\bar{q}_6$, $\bar{q}_8$  of the structures found in the system without walls, $L_x=L_y=20$, $L_z=19$}
\label{tab:found_param}
}
\end{table}

\begin{table}[ht]
\center{
\begin{tabular}{|c|c||c|c|c|}
\hline
 & color & $\bar{q}_4$ & $\bar{q}_6$ & $\bar{q}_8$\\
\hline
\hline
Crystal 1 & black & $0.151 \pm 0.036$ & $0.504 \pm 0.018$ & $0.300 \pm 0.038$ \\
\hline
Crystal 2 & red & $0.211 \pm 0.047$ & $0.552 \pm 0.019$ & $0.360 \pm 0.031$ \\
\hline
Melt & green & $0.225 \pm 0.073$ & $0.385 \pm 0.071$ & $0.330 \pm 0.058$\\
\hline

\end{tabular}
    \caption{parameters $\bar{q}_4$, $\bar{q}_6$, $\bar{q}_8$  of the structures found in the system with repulsive walls, $L_x=L_y=20$, $L_z=20$}
\label{tab:found_param_walls}
}
\end{table}

The method was also tested on a system with repulsive walls ( Tab. \ref{tab:found_param_walls}). 
In these two studied systems, structures with identical sets of $\bar{q}_j^k$ are formed. 
If we compare the spatial arrangement of the structures of Tab. \ref{tab:number_bulk2} and Tab. \ref{tab:number_walls2} (see Supplementary materials below), the dominant crystalline phase in both systems is the $Crystal\;2$ (red). 
In a system with repulsive walls, the alternation of crystal phases is observed parallel to the walls, while in a system without walls, alternation occurs in the planes located at an angle to the planes of the simulation box.
The conformations are dominated by layered crystal phases, with the $Crystal\;2$ dominating in the vicinity of the walls.

The received data Tab. \ref{tab:found_param} it is said that the structures found do not exactly define any of the test structures ($sc$, $bcc$, $fcc$, $hcp$). 
However, taking into account the estimated spread of the values obtained, it can be seen that the parameters of the $Crystal\;2$ (red structure) are closest to $fcc$, while the $Crystal\;1$ (black structure) is close in parameters to $hcp$. 
The only significant difference is observed for the parameter $q_4$ of the $Crystal\;1$ and $hcp$. 
The deviation of this parameter from the reference value indicates the deformation of the structure. 
Indeed, when considering the projections of the centers of the particles of the system on different planes, irregular, deformed hexagons were observed. 
The reason for the deformation is the connectivity in the chain and the dimensions of the box. So, for example, if in low energy chains are stretched along the $y$ axis one after the other, then in the perpendicular plane $XZ$ we see sections of only 360 chains. 
By no such transformations, it is impossible to stretch particles stacked in a regular polygon on the plane to fit it into a rectangle of size $L_x \cdot L_z = 20 \cdot 19$. 
Note that although in reality the $y$ direction is not highlighted, the final, small size of the system has a strong influence on the laying in our case. 
Nevertheless, the organization of short chains into layers was tested on a similar system in the work of T. Shakirov\cite{tim_model1}.

Thus, it can be argued that the approaches developed in this work are valid within the framework of this model.

  \section{Conclusions}
\label{sec:concl}

We have suggested two new procedures of analysis of noisy crystalline structures which appear in computer simulations of soft matter systems. 

The first of these two procedures -- a noise reduction -- is a special averaging over neighbors of a chosen particle before calculation of bond local order parameters (Steinhardt parameters). This procedure leads to an essential reduction of noise in particles' positions and allows to determine the local crystalline symmetry more reliably. 

Another procedure -- lattice reconstruction -- allows to reconstruct the ``ideal'' lattice structure in the whole simulation box that would be most close to a real noisy crystalline symmetry when it is first detected locally and then averaged over the whole box.

We plan to apply both these procedures to analysis of crystallization transitions in the melts and in thin films of short semiflexible chains of tangent hard spheres. 

  \setcounter{secnumdepth}{0}

  \newpage
  \interlinepenalty=10000
  
  \bibliography{resources/library}

\begin{thebibliography}{33}
\providecommand{\natexlab}[1]{#1}
\providecommand{\url}[1]{\texttt{#1}}
\expandafter\ifx\csname urlstyle\endcsname\relax
  \providecommand{\doi}[1]{doi: #1}\else
  \providecommand{\doi}{doi: \begingroup \urlstyle{rm}\Url}\fi

\bibitem[Brown et~al.(2018)Brown, Shah, Morrell, Zubich, Wagner, Denton, and Hobbie]{Brown2018}
S.~L. Brown, V.~D. Shah, M.~V. Morrell, M.~Zubich, A.~Wagner, A.~R. Denton, and E.~K. Hobbie.
\newblock Superlattice formation in colloidal nanocrystal suspensions: Hard-sphere freezing and depletion effects.
\newblock \emph{Physical Review E}, 98\penalty0 (6):\penalty0 062616, 2018.
\newblock \doi{10.1103/PhysRevE.98.062616}.

\bibitem[Escobedo and de~Pablo(1997)]{dePablo1997}
F.~A. Escobedo and J.~J. de~Pablo.
\newblock {Monte Carlo simulation of athermal mesogenic chains: Pure systems, mixtures, and constrained environments}.
\newblock \emph{The Journal of Chemical Physics}, 106\penalty0 (23):\penalty0 9858--9868, 1997.
\newblock \doi{10.1063/1.473874}.

\bibitem[Schmidt and Löwen(1997)]{Loewen1997}
M.~Schmidt and H.~Löwen.
\newblock Phase diagram of hard spheres confined between two parallel plates.
\newblock \emph{Physical Review E}, 55\penalty0 (6):\penalty0 7228, 1997.
\newblock \doi{10.1103/PhysRevE.55.7228}.

\bibitem[Zykova-Timan et~al.(2010)Zykova-Timan, Horbach, and Binder]{Horbach2010}
T.~Zykova-Timan, J.~Horbach, and K.~Binder.
\newblock {Monte Carlo simulations of the solid-liquid transition in hard spheres and colloid-polymer mixtures}.
\newblock \emph{The Journal of Chemical Physics}, 133\penalty0 (1):\penalty0 014705, 2010.
\newblock \doi{10.1063/1.3455504}.

\bibitem[Karayiannis and Laso(2008)]{laso_ths1}
N.~Ch. Karayiannis and M.~Laso.
\newblock Dense and nearly jammed random packings of freely jointed chains of tangent hard spheres.
\newblock \emph{Physical Review Letters}, 100\penalty0 (5):\penalty0 050602, 2008.
\newblock \doi{10.1103/PhysRevLett.100.050602}.

\bibitem[Karayiannis et~al.(2009{\natexlab{a}})Karayiannis, Foteinopoulou, and Laso]{laso_ths2}
N.~Ch. Karayiannis, K.~Foteinopoulou, and M.~Laso.
\newblock The structure of random packings of freely jointed chains of tangent hard spheres.
\newblock \emph{The Journal of Chemical Physics}, 130\penalty0 (16):\penalty0 164908, 2009{\natexlab{a}}.
\newblock \doi{10.1063/1.3117903}.

\bibitem[Karayiannis et~al.(2009{\natexlab{b}})Karayiannis, Foteinopoulou, and Laso]{LasoPRL2009}
N.~Ch. Karayiannis, K.~Foteinopoulou, and M.~Laso.
\newblock Entropy-driven crystallization in dense systems of athermal chain molecules.
\newblock \emph{Physical Review Letters}, 103\penalty0 (4):\penalty0 045703, 2009{\natexlab{b}}.
\newblock \doi{10.1103/PhysRevLett.103.045703}.

\bibitem[Karayiannis et~al.(2010)Karayiannis, Foteinopoulou, Abrams, and Laso]{LasoSoftMat2010}
N.~Ch. Karayiannis, K.~Foteinopoulou, C.~F. Abrams, and M.~Laso.
\newblock Modeling of crystal nucleation and growth in athermal polymers: Self-assembly of layered nano-morphologies.
\newblock \emph{Soft Matter}, 6:\penalty0 2160--2173, 2010.
\newblock \doi{10.1039/b923369e}.

\bibitem[Karayiannis et~al.(2012)Karayiannis, Foteinopoulou, and Laso]{laso_ths3}
N.~Ch. Karayiannis, K.~Foteinopoulou, and M.~Laso.
\newblock Spontaneous crystallization in athermal polymer packings.
\newblock \emph{International journal of molecular sciences}, 14\penalty0 (1):\penalty0 332--358, 2012.
\newblock \doi{10.3390/ijms14010332}.

\bibitem[Karayiannis et~al.(2013)Karayiannis, Foteinopoulou, and Laso]{LazoPhilMag2013}
N.~Ch. Karayiannis, K.~Foteinopoulou, and M.~Laso.
\newblock Jamming and crystallization in athermal polymer packings.
\newblock \emph{Philosophical Magazine}, 93\penalty0 (31-33):\penalty0 4108--4131, 2013.
\newblock \doi{10.1080/14786435.2013.815377}.

\bibitem[Cochran and Chiew(2006)]{rdf-FJTHS}
T.~W. Cochran and Y.~C. Chiew.
\newblock Radial distribution function of freely jointed hard-sphere chains in the solid phase.
\newblock \emph{The Journal of Chemical Physics}, 124\penalty0 (7):\penalty0 074901, 2006.
\newblock \doi{10.1063/1.2167644}.

\bibitem[Rycroft(2009{\natexlab{a}})]{voro}
Chris~H. Rycroft.
\newblock {VORO++: A three-dimensional Voronoi cell library in C++}.
\newblock \emph{Chaos}, 19\penalty0 (4):\penalty0 041111, 2009{\natexlab{a}}.
\newblock \doi{10.1063/1.3215722}.

\bibitem[Rycroft(2009{\natexlab{b}})]{voro_manual}
C.~H. Rycroft.
\newblock Voro++: a three-dimensional {Voronoi} cell library in {C++}.
\newblock https://math.lbl.gov/voro++/doc, 2009{\natexlab{b}}.

\bibitem[ans H.C.~Andersen(1987)]{cna}
J.D.~Honeycutt ans H.C.~Andersen.
\newblock Molecular dynamics study of melting and freezing of small lennard-jones clusters.
\newblock \emph{J. Phys. Chem}, 31\penalty0 (19):\penalty0 4950--4963, 1987.
\newblock \doi{https://pubs.acs.org/doi/pdf/10.1021/j100303a014}.

\bibitem[Faken and Jónsson(1994)]{cna2}
D.~Faken and H.~Jónsson.
\newblock Systematic analysis of local atomic structure combined with {3D} computer graphics.
\newblock \emph{Computational Materials Science}, 2\penalty0 (2):\penalty0 279--286, 1994.
\newblock \doi{10.1016/0927-0256(94)90109-0}.

\bibitem[Stukowski(2012)]{acna}
A.~Stukowski.
\newblock Structure identification methods for atomistic simulations of crystalline materials.
\newblock \emph{Modelling and Simulation in Materials Science and Engineering}, 20\penalty0 (4):\penalty0 045021, 2012.
\newblock \doi{10.1088/0965-0393/20/4/045021}.

\bibitem[Larsen(2020)]{icna}
P.~M. Larsen.
\newblock Revisiting the common neighbour analysis and the centrosymmetry parameter.
\newblock 2020.
\newblock \doi{10.48550/arXiv.2003.08879}.

\bibitem[Anwar et~al.(2013)Anwar, Turci, and Schilling]{schilling}
M.~Anwar, F.~Turci, and T.~Schilling.
\newblock Crystallization mechanism in melts of short n-alkane chains.
\newblock \emph{The Journal of Chemical Physics}, 139\penalty0 (21):\penalty0 214904, 2013.
\newblock \doi{10.1063/1.4835015}.

\bibitem[Kos et~al.(2021)Kos, Ivanov, and Chertovich]{kos}
P.~I. Kos, V.~A. Ivanov, and A.~V. Chertovich.
\newblock Crystallization of semiflexible polymers in melts and solutions.
\newblock \emph{Soft Matter}, 17\penalty0 (9):\penalty0 2392--2403, 2021.
\newblock \doi{10.1039/D0SM01545H}.

\bibitem[Steinhardt et~al.(1983)Steinhardt, Nelson, and Ronchetti]{steinhardt_main}
P.~J. Steinhardt, D.~R. Nelson, and M.~Ronchetti.
\newblock Bond-orientational order in liquids and glasses.
\newblock \emph{Phys. Rev. B}, 28\penalty0 (2):\penalty0 784--805, 1983.
\newblock \doi{https://doi.org/10.1103/PhysRevB.28.784}.

\bibitem[Lechner and Dellago(2008)]{dellago1}
W.~Lechner and C.~Dellago.
\newblock Accurate determination of crystal structures based on averaged local bond order parameters.
\newblock \emph{J. Chem. Phys}, 128:\penalty0 114707, 2008.
\newblock \doi{https://doi.org/10.1063/1.2977970}.

\bibitem[Mickel et~al.(2013)Mickel, Kapfer, Schröder-Turk, and Mecke]{Mecke2013}
W.~Mickel, S.~C. Kapfer, G.~E. Schröder-Turk, and K.~Mecke.
\newblock Shortcomings of the bond orientational order parameters for the analysis of disordered particulate matter.
\newblock \emph{The Journal of Chemical Physics}, 138\penalty0 (4):\penalty0 044501, 2013.
\newblock \doi{10.1063/1.4774084}.

\bibitem[Eslami et~al.(2018)Eslami, Sedaghat, and Müller-Plathe]{eslami}
H.~Eslami, P.~Sedaghat, and F.~Müller-Plathe.
\newblock Local bond order parameters for accurate determination of crystal structures in two and three dimensions.
\newblock \emph{Phys. Chem. Chem. Phys}, 20:\penalty0 27059--27068, 2018.
\newblock \doi{10.1039/C8CP05248D}.

\bibitem[Haeberle et~al.(2019)Haeberle, Sperl, and Born]{Haeberle2019}
J.~Haeberle, M.~Sperl, and P.~Born.
\newblock Distinguishing noisy crystalline structures using bond orientational order parameters.
\newblock \emph{The European Physical Journal E}, 42\penalty0 (149):\penalty0 1--7, 2019.
\newblock \doi{10.1140/epje/i2019-11915-7}.

\bibitem[Karayiannis et~al.(2009{\natexlab{c}})Karayiannis, Foteinopoulou, and Laso]{LasoCCE2009}
N.~Ch. Karayiannis, K.~Foteinopoulou, and M.~Laso.
\newblock The characteristic crystallographic element norm: A descriptor of local structure in atomistic and particulate systems.
\newblock \emph{The Journal of Chemical Physics}, 130\penalty0 (7):\penalty0 074704, 2009{\natexlab{c}}.
\newblock \doi{10.1063/1.3077294}.

\bibitem[Ramos et~al.(2020)Ramos, Herranz, Foteinopoulou, Karayiannis, and Laso]{LasoCryst2020}
P.~M. Ramos, M.~Herranz, K.~Foteinopoulou, N.~Ch. Karayiannis, and M.~Laso.
\newblock Identification of local structure in 2-d and 3-d atomic systems through crystallographic analysis.
\newblock \emph{Crystals}, 10\penalty0 (11):\penalty0 1008, 2020.
\newblock \doi{10.3390/cryst10111008}.

\bibitem[Martínez-Fernández et~al.(2023)Martínez-Fernández, Herranz, Foteinopoulou, Karayiannis, and Laso]{LasoPolym2023-1}
D.~Martínez-Fernández, M.~Herranz, K.~Foteinopoulou, N.~Ch. Karayiannis, and M.~Laso.
\newblock Local and global order in dense packings of semi-flexible polymers of hard spheres.
\newblock \emph{Polymers}, 15\penalty0 (3):\penalty0 551, 2023.
\newblock \doi{10.3390/polym15030551}.

\bibitem[Ramos et~al.(2021)Ramos, Herranz, Foteinopoulou, Karayiannis, and Laso]{LasoPolym2021}
P.~M. Ramos, M.~Herranz, K.~Foteinopoulou, N.~Ch. Karayiannis, and M.~Laso.
\newblock Entropy-driven heterogeneous crystallization of hard-sphere chains under unidimensional confinement.
\newblock \emph{Polymers}, 13\penalty0 (9):\penalty0 1352, 2021.
\newblock \doi{10.3390/polym13091352}.

\bibitem[Janke and Paul(2016)]{samc_paul_janke}
W.~Janke and W.~Paul.
\newblock Thermodynamics and structure of macromolecules from flat-histogram {Monte Carlo} simulations.
\newblock \emph{Soft Matter}, 12:\penalty0 642--657, 2016.
\newblock \doi{10.1039/C5SM01919B}.

\bibitem[Shakirov and Paul(2018)]{tim_model1}
T.~Shakirov and W.~Paul.
\newblock Crystallization in melts of short, semiflexible hard polymer chains: An interplay of entropies and dimensions.
\newblock \emph{Phys. Rev. E}, 97\penalty0 (4):\penalty0 042501, 2018.
\newblock \doi{https://doi.org/10.1103/PhysRevE.97.042501}.

\bibitem[Shakirov(2019)]{tim_model2}
T.~Shakirov.
\newblock Crystallization in melts of short, semi-flexible hard-sphere polymer chains: The role of the non-bonded interaction range.
\newblock \emph{Entropy}, 21\penalty0 (9):\penalty0 856, 2019.
\newblock \doi{https://doi.org/10.3390/e21090856}.

\bibitem[Anderberg(1973)]{kmeans}
M.R. Anderberg.
\newblock \emph{Cluster analysis for applications: probability and mathematical statistics: a series of monographs and textbooks}.
\newblock Academic Press, New York, 1973.

\bibitem[Likas et~al.(2003)Likas, Vlassis, and Verbeek]{kmeans_var}
A.~Likas, N.~Vlassis, and J.J. Verbeek.
\newblock The global k-means clustering algorithm.
\newblock \emph{Pattern Recognition}, 36\penalty0 (2):\penalty0 451--461, 2003.
\newblock \doi{https://doi.org/10.1016/S0031-3203(02)00060-2}.

\end{thebibliography}
    \bibliographystyle{unsrtnat} 

\clearpage

\newpage

\newgeometry{twoside=false, left=2cm, right=1.5cm, top=2cm, bottom=2cm}

\onecolumngrid
\section{Supplementary}
\label{sec:supl}
\subsection{Local bond order parameters}
\label{subsec:local_supl}

\begin{table}[h]
\center{
\begin{tabular}{|c||c|c|c|c|}
\hline
 & $sc$ & $bcc$ & $fcc$ & $hcp$\\
\hline
\hline
Neighbors & 6 & 14 & 12 & 12 \\
\hline
$q_2$ & 0.0 & 0.0 & 0.0 & 0.0 \\
\hline
$q_3$ & 0.0 & 0.0 & 0.0 & 0.076073 \\
\hline
$q_4$ & 0.763763 & 0.036370 & 0.190941 & 0.097222 \\
\hline
$q_5$ & 0.0 & 0.0 & 0.0 & 0.251586 \\
\hline
$q_6$ & 0.353553 & 0.510688 & 0.574524 & 0.484762 \\
\hline
$q_7$ & 0.0 & 0.0 & 0.0 & 0.310815 \\
\hline
$q_8$ & 0.718070 & 0.429322 & 0.403915 & 0.316992 \\
\hline
$q_9$ & 0.0 & 0.0 & 0.0 & 0.137851 \\
\hline
$q_{10}$ & 0.411425 & 0.195191 & 0.012857 & 0.010169 \\
\hline
$w_2$ & 0.0 & 0.0 & 0.0 & 0.0 \\
\hline
$w_3$ & 0.0 & 0.0 & 0.0 & 0.0 \\
\hline
$w_4$ & 0.159317 & 0.159317 & -0.159317 & 0.134097 \\
\hline
$w_5$ & 0.0 & 0.0 & 0.0 & 0.0 \\
\hline
$w_6$ & 0.013161 & 0.013161 & -0.013161 & -0.012442 \\
\hline
$w_7$ & 0.0 & 0.0 & 0.0 & 0.0 \\
\hline
$w_8$ & 0.058455 & 0.058455 & 0.058455 & 0.051259 \\
\hline
$w_9$ & 0.0 & 0.0 & 0.0 & 0.0 \\
\hline
$w_{10}$ & 0.090130 & -0.090130 & -0.090130 & -0.079851 \\

\hline

\end{tabular}
    \caption{local bond order parameters $q_2$ - $q_{10}$, $w_2$ - $w_{10}$ for perfect $sc$, $bcc$, $fcc$, $hcp$ structures.}
\label{tab:bop_noise}
}
\end{table}

\newpage

\begin{table}[H]
\center{
\begin{tabular}{| c|| m{1.8cm}| m{1.8cm}| m{1.8cm}| m{4cm}|}
\hline
$E$ & Crystal 1 (black) & Crystal 2 (red) & Melt (green) & View\\
\hline
\hline
-5727 & 2803 (39\%) & 4070 (56\%) & 327 (5\%) & 
    {\center{
    \begin{minipage}[t]{.295\textwidth}
      \includegraphics[width=0.85\linewidth, height=30mm]{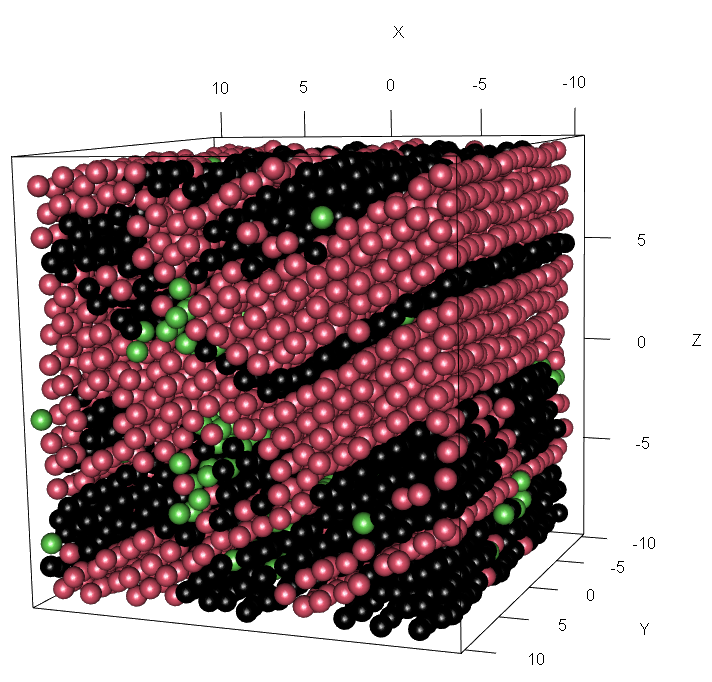}
    \end{minipage}
    }} \\ 
\hline
-5104 & 2449 (34\%) & 3329 (46\%) & 1422 (20\%) & 
    {\center{\begin{minipage}[t]{.295\textwidth}
      \includegraphics[width=0.85\linewidth, height=30mm]{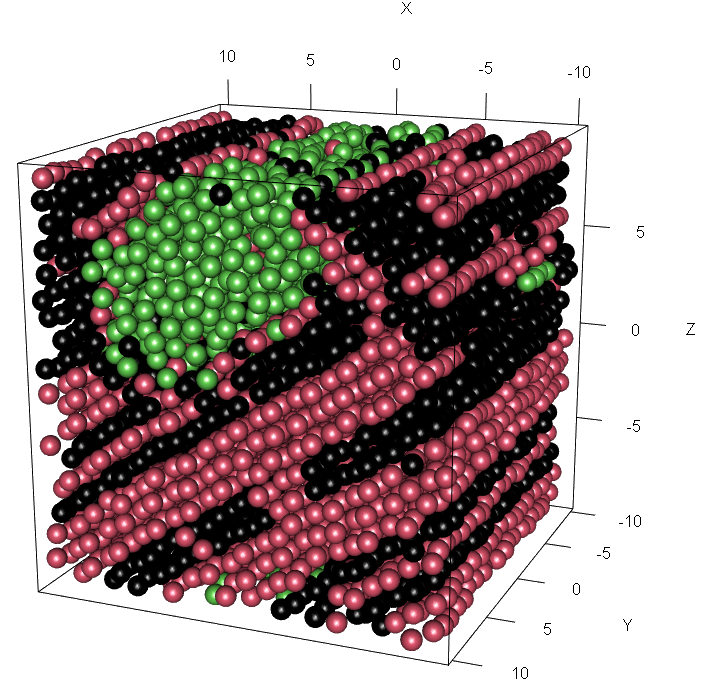}
    \end{minipage}
    }} \\ 
\hline

-4715 & 1601 (22\%) & 3194 (44\%) & 2405 (34\%) & 
    {\center{\begin{minipage}[t]{.295\textwidth}
      \includegraphics[width=0.85\linewidth, height=30mm]{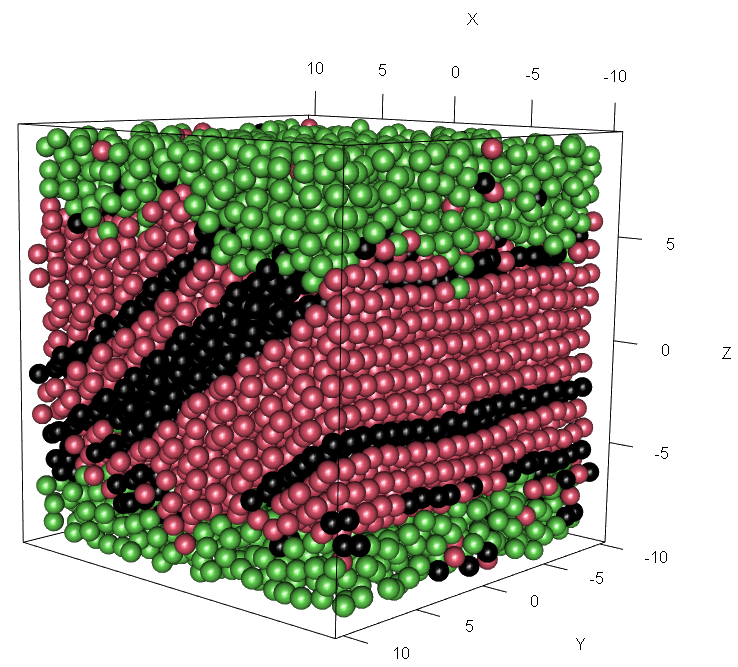}
    \end{minipage}
    }} \\ 
\hline

-3408 & 1034 (14\%) & 1266 (18\%) & 4900 (68\%) & 
    {\center{\begin{minipage}[t]{.295\textwidth}
      \includegraphics[width=0.85\linewidth, height=30mm]{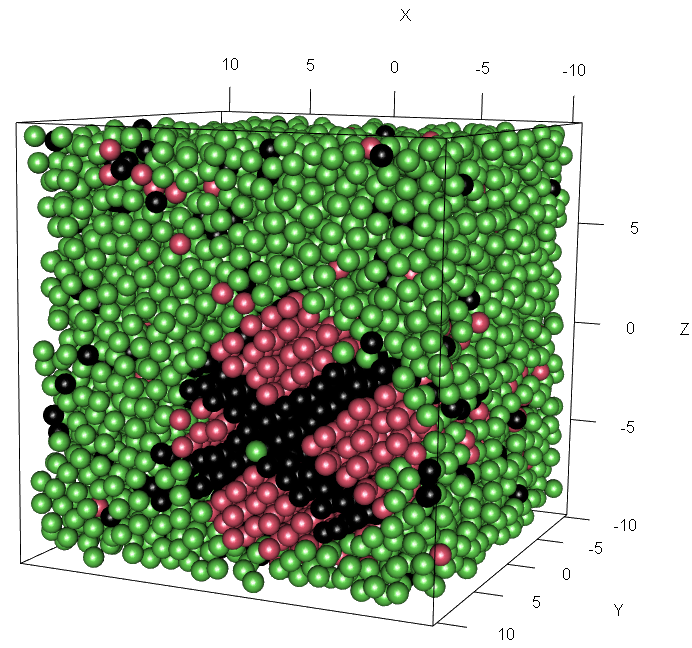}
    \end{minipage}
    }} \\ 
\hline

-2040 & 394 (5\%) & 113 (2\%) & 6693 (93\%) & 
    {\center{\begin{minipage}[t]{.295\textwidth}
      \includegraphics[width=0.85\linewidth, height=30mm]{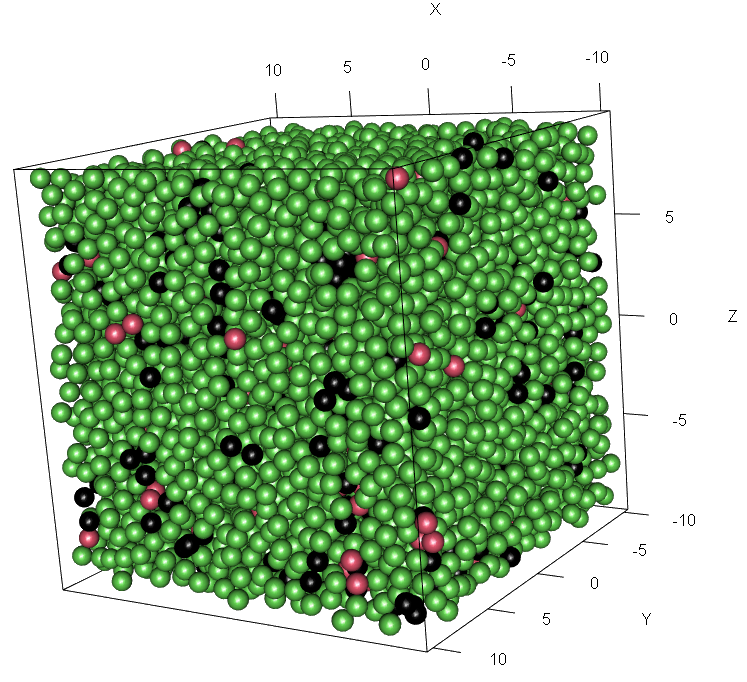}
    \end{minipage}
    }} \\ 
\hline

\hline
    
\end{tabular}
    \caption{The number of particles (and percentage) related to a clusters for different energies in the system without walls, $L_x=L_y=20$, $L_z=19$}
\label{tab:number_bulk2}
}
\end{table}

\begin{table}[H]
\center{
\begin{tabular}{| c|| m{1.8cm}| m{1.8cm}| m{1.8cm}| m{4cm}|}
\hline
$E$ & Crystal 1 (black) & Crystal 2 (red) & Melt (green) & View\\
\hline
\hline
-5754 & 2316 (32\%) & 4587 (64\%) & 297 (4\%) & 
    {\center{
    \begin{minipage}[t]{.295\textwidth}
      \includegraphics[width=0.85\linewidth, height=30mm]{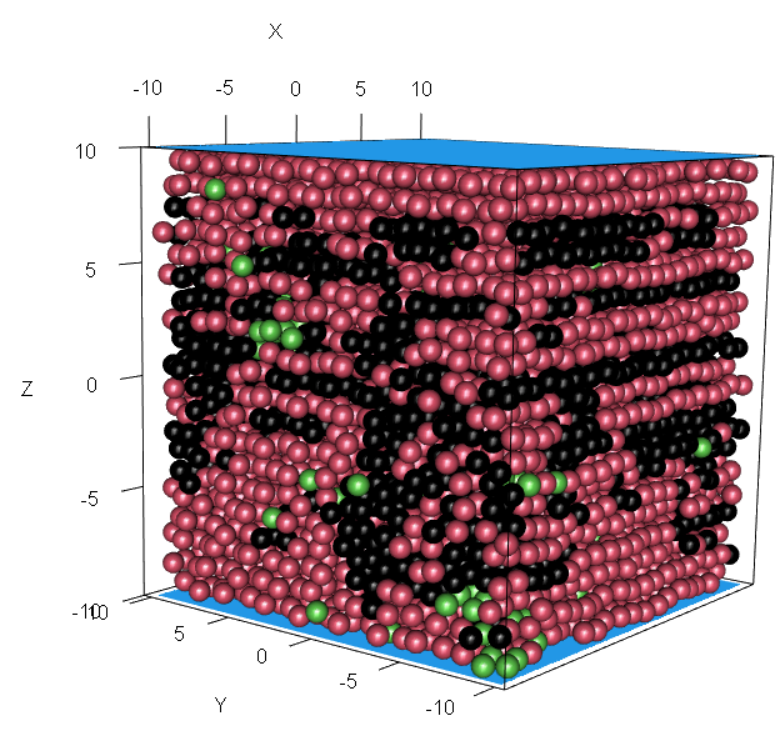}
    \end{minipage}
    }} \\ 
\hline

-5205 & 1262 (17\%) & 4289 (60\%) & 1649 (23\%) & 
    {\center{\begin{minipage}[t]{.295\textwidth}
      \includegraphics[width=0.85\linewidth, height=30mm]{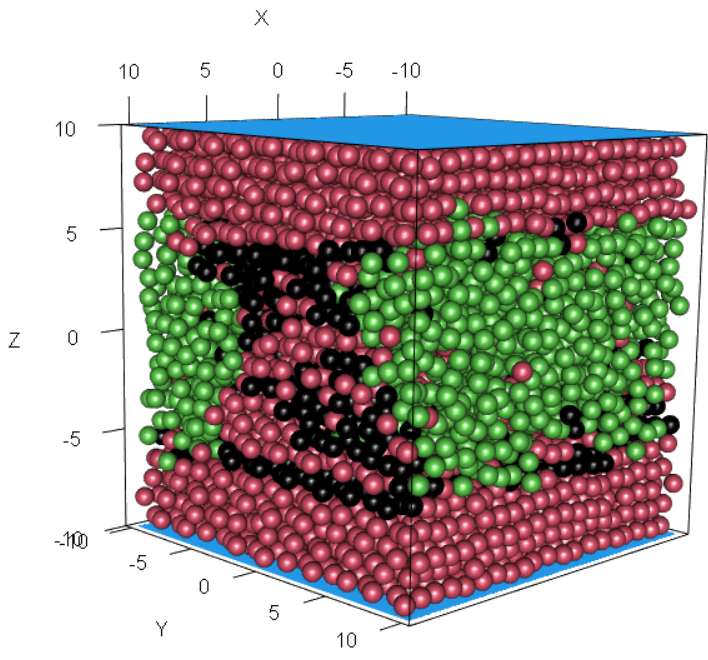}
    \end{minipage}
    }} \\ 
\hline

-4740 & 1626 (23\%) & 2548 (35\%) & 3026 (42\%) & 
    {\center{\begin{minipage}[t]{.295\textwidth}
      \includegraphics[width=0.85\linewidth, height=30mm]{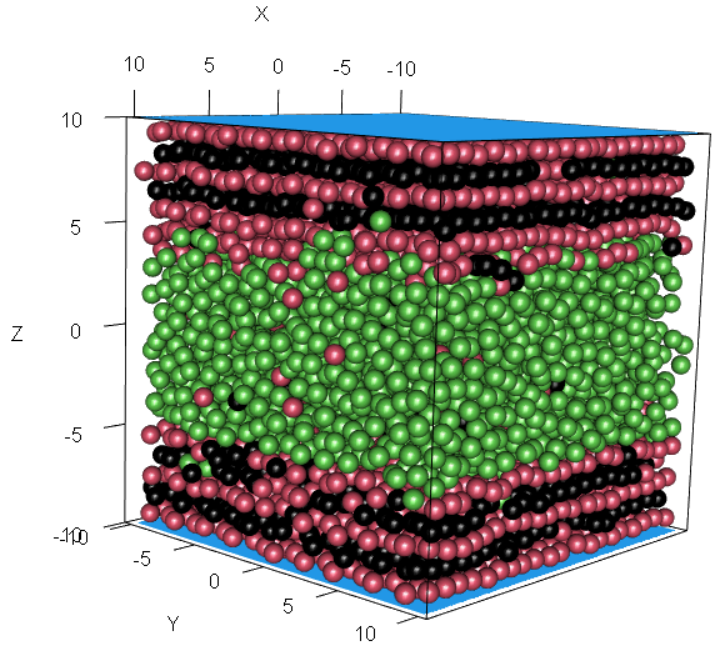}
    \end{minipage}
    }} \\ 
\hline

-3600 & 282 (4\%) & 1728 (24\%) & 5190 (72\%) & 
    {\center{\begin{minipage}[t]{.295\textwidth}
      \includegraphics[width=0.85\linewidth, height=30mm]{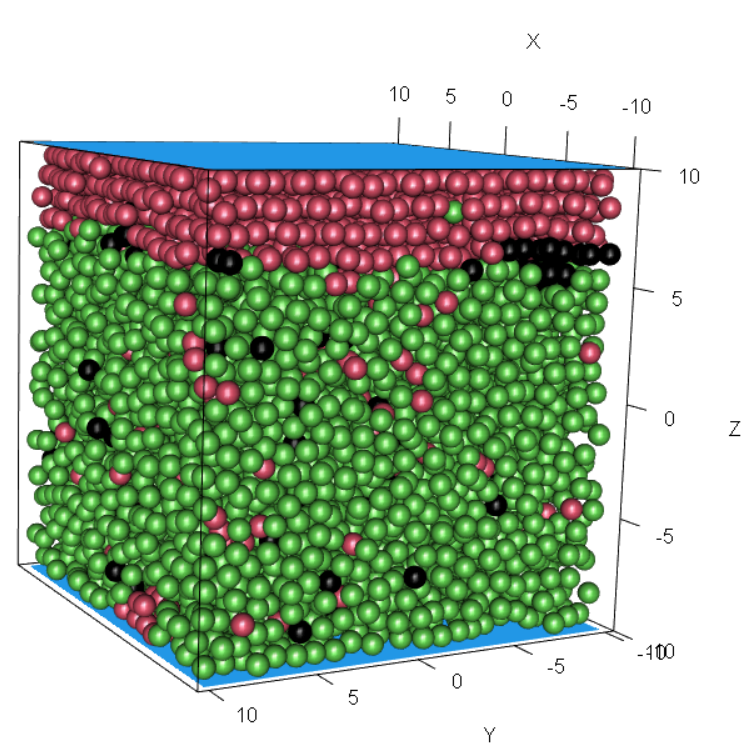}
    \end{minipage}
    }} \\ 
\hline

-2050 & 237 (3\%) & 475 (7\%) & 6488 (90\%) & 
    {\center{\begin{minipage}[t]{.295\textwidth}
      \includegraphics[width=0.85\linewidth, height=30mm]{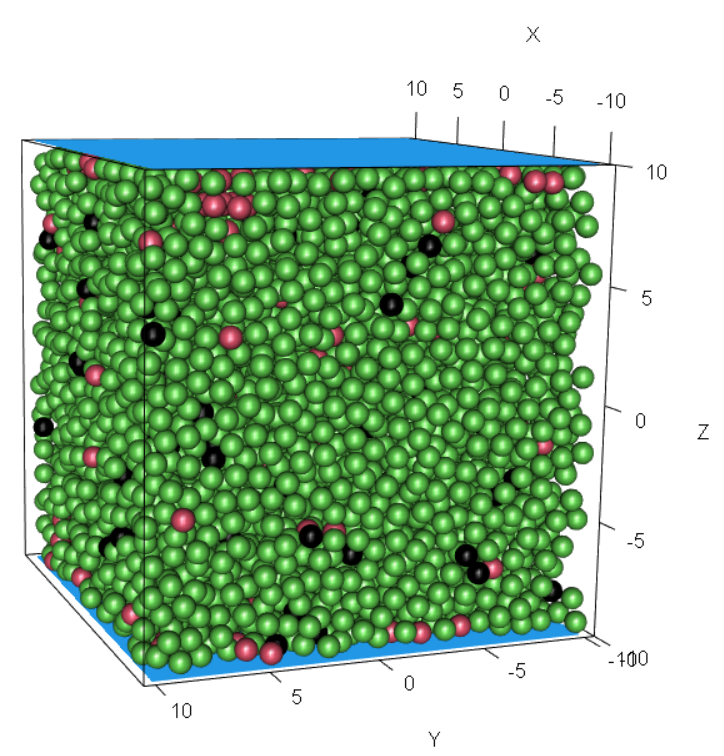}
    \end{minipage}
    }} \\ 
\hline

\hline
    
\end{tabular}
    \caption{The number of particles (and percentage) related to a clusters for different energies in the system with repulsive walls, $L_x=L_y=20$, $L_z=20$}
\label{tab:number_walls2}
}
\end{table}

\newpage
\subsubsection{P. J. Steinhardt, R. Nelson and M. Ronchetti parameters}
\label{subsubsec:steinhardt_supl}

\begin{figure}[H]
\center{
    \begin{minipage}[h]{0.4\linewidth}
\center{\includegraphics[width=1\linewidth]{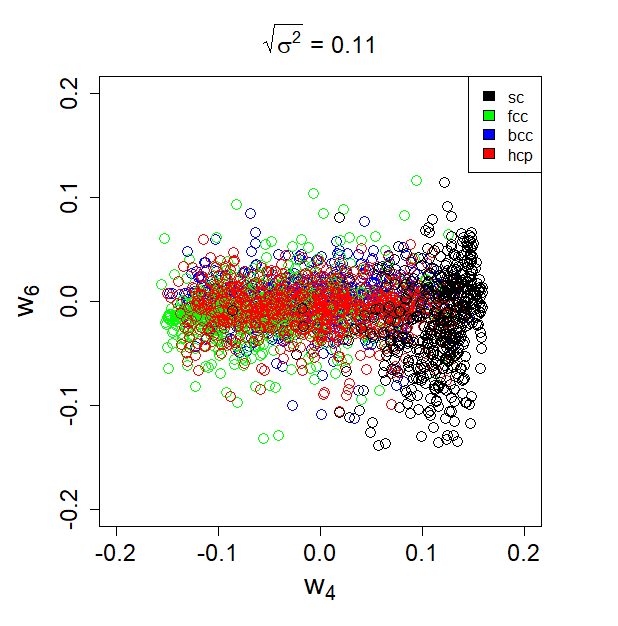} \\ (a)}
\end{minipage}
\begin{minipage}[h]{0.4\linewidth}
\center{\includegraphics[width=1\linewidth]{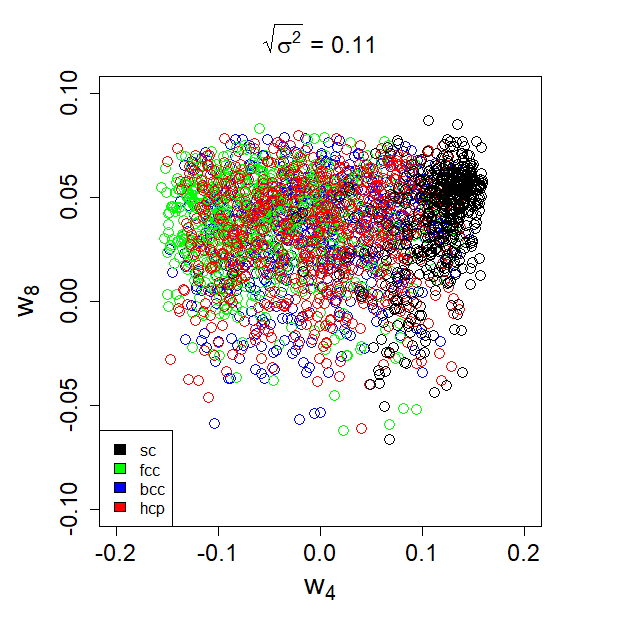} \\ (b)}
\end{minipage}
\vfill
\begin{minipage}[h]{0.4\linewidth}
\center{\includegraphics[width=1\linewidth]{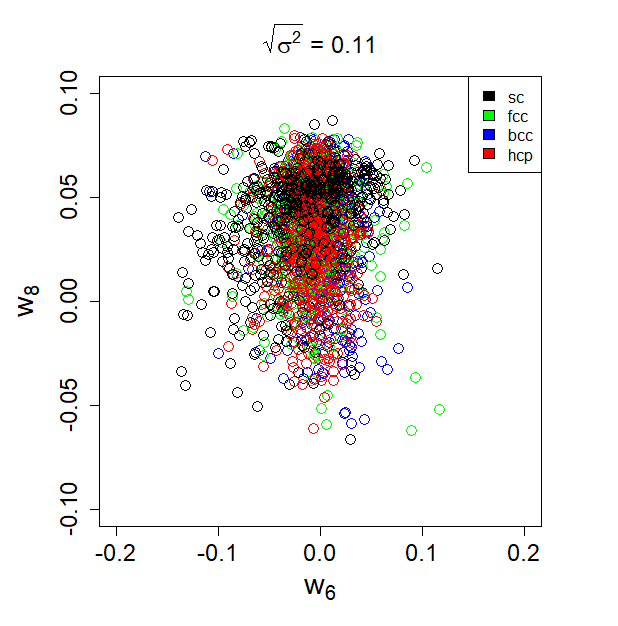} \\ (c)}
\end{minipage}
    \begin{minipage}[h]{0.4\linewidth}
\center{\includegraphics[width=1\linewidth]{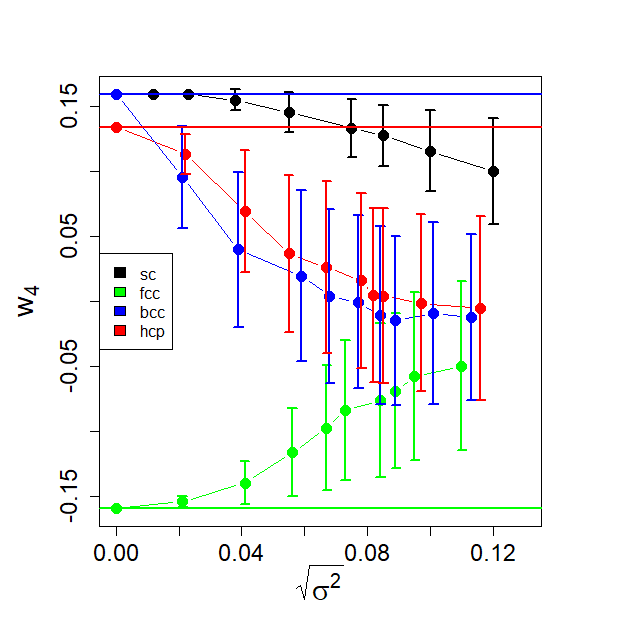} \\ (d)}
\end{minipage}
\vfill
    \begin{minipage}[h]{0.4\linewidth}
\center{\includegraphics[width=1\linewidth]{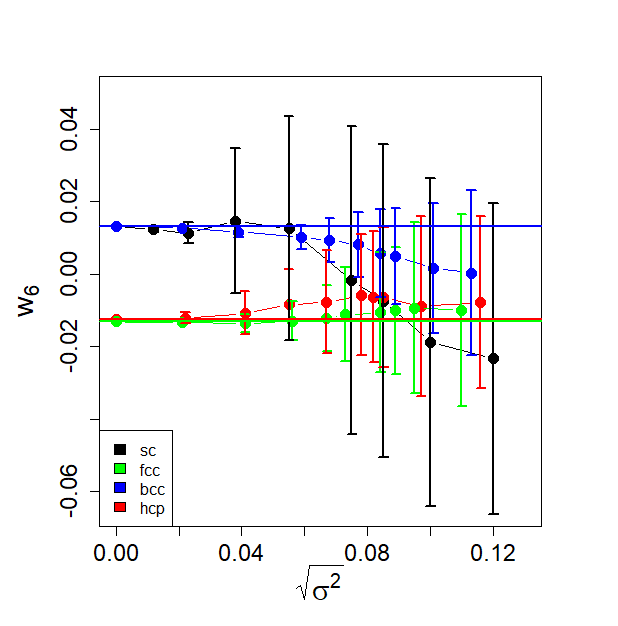} \\ (e)}
\end{minipage}
    \begin{minipage}[h]{0.4\linewidth}
\center{\includegraphics[width=1\linewidth]{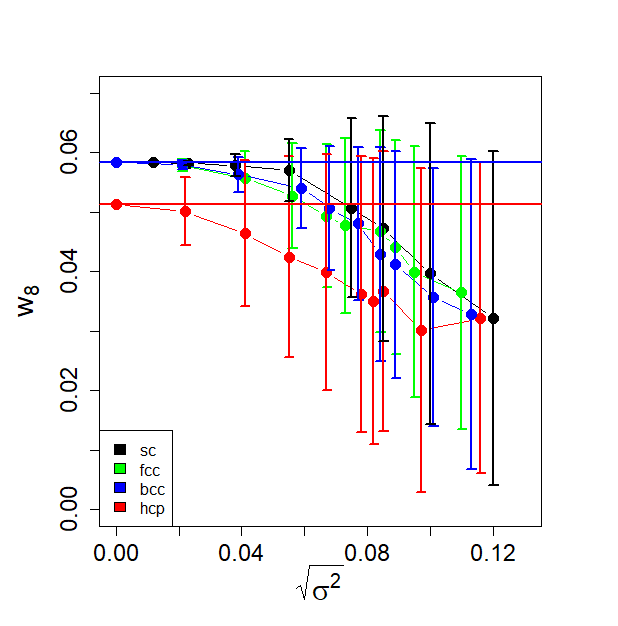} \\ (f)}
\end{minipage}
    \caption{Local bond order parameters $w_4-w_6$, $w_4-w_8$, $w_6-w_8$ planes (a-c) and mean values of $w_4$, $w_6$, $w_8$ (d-f) for test structures.}
    \label{fig:test_w_comp1}
    }
\end{figure}

\newpage
\subsubsection{W. Lechner and C. Dellago parameters}
\label{subsubsec:dellago_supl}

\begin{figure}[H]
\center{
    \begin{minipage}[h]{0.4\linewidth}
\center{\includegraphics[width=1\linewidth]{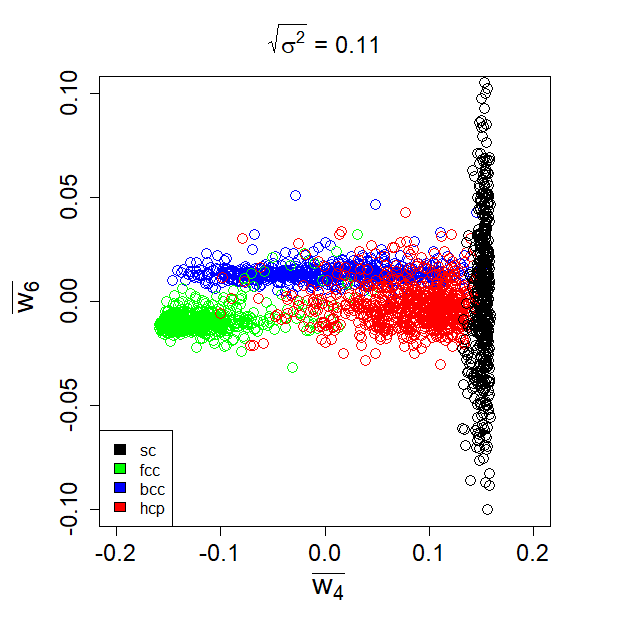} \\ (a)}
\end{minipage}
\begin{minipage}[h]{0.4\linewidth}
\center{\includegraphics[width=1\linewidth]{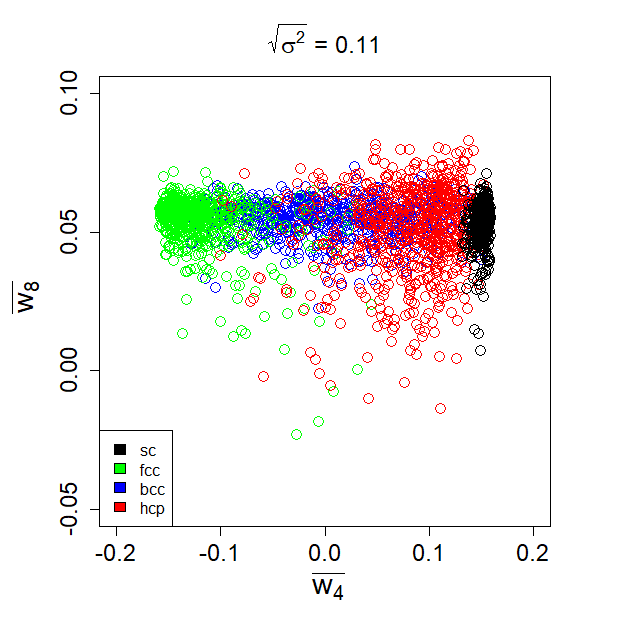} \\ (b)}
\end{minipage}
\vfill
\begin{minipage}[h]{0.4\linewidth}
\center{\includegraphics[width=1\linewidth]{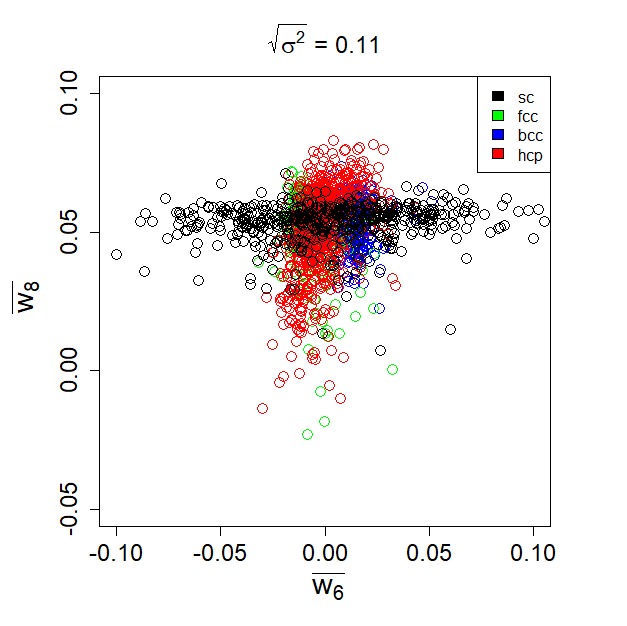} \\ (c)}
\end{minipage}
    \begin{minipage}[h]{0.4\linewidth}
\center{\includegraphics[width=1\linewidth]{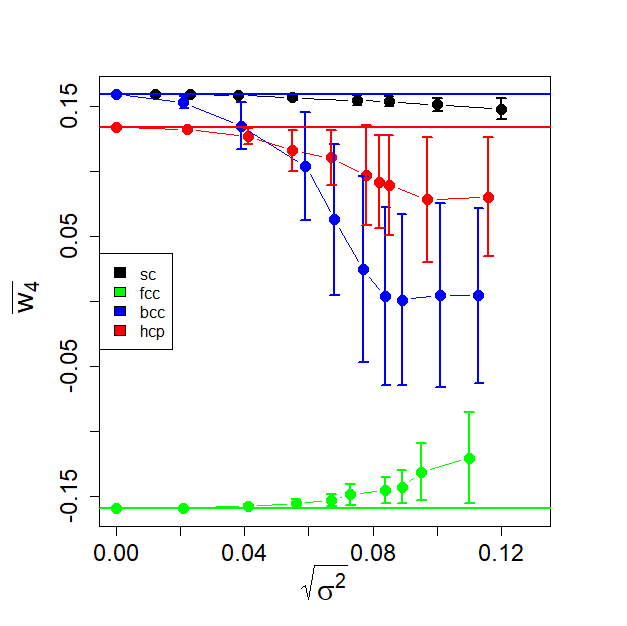} \\ (d)}
\end{minipage}
\vfill
    \begin{minipage}[h]{0.4\linewidth}
\center{\includegraphics[width=1\linewidth]{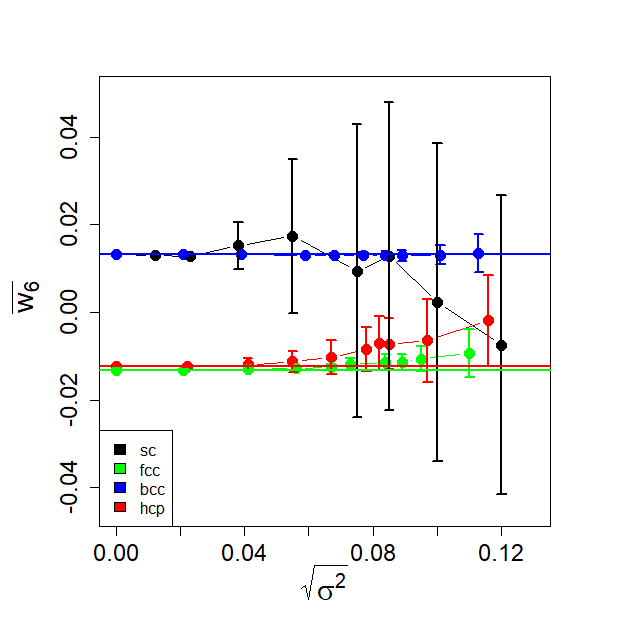} \\ (e)}
\end{minipage}
    \begin{minipage}[h]{0.4\linewidth}
\center{\includegraphics[width=1\linewidth]{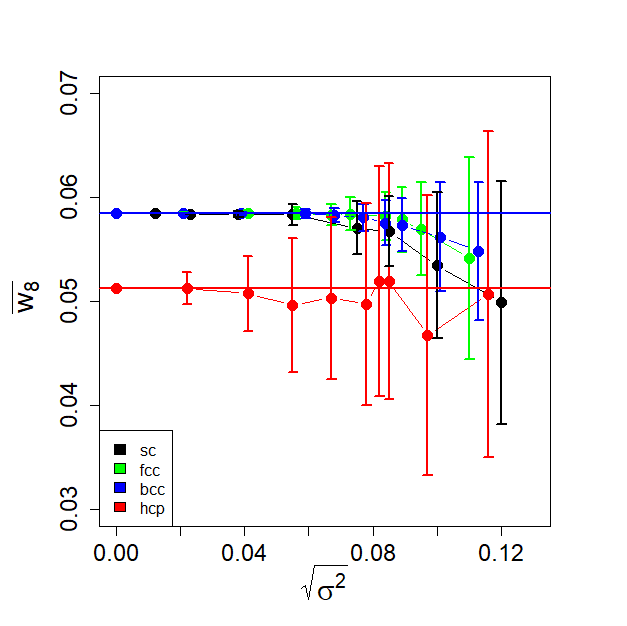} \\ (f)}
\end{minipage}
    \caption{Average local bond order parameters $\bar{w}_4-\bar{w}_6$, $\bar{w}_4-\bar{w}_8$, $\bar{w}_6-\bar{w}_8$ planes (a-c) and mean values of $\bar{w}_4$, $\bar{w}_6$, $\bar{w}_8$ (d-f) for test structures.}
    \label{fig:test_w_comp2}
    }
\end{figure}

\newpage
\subsection{Noise reduction procedure}
\label{subsec:noise_reduction_supl}

\begin{figure}[H]
\center{
    \begin{minipage}[h]{0.40\linewidth}
\center{\includegraphics[width=1\linewidth]{./resources/figures/noise_reduction/points/q46.png} \\ (a)}
\end{minipage}
\begin{minipage}[h]{0.40\linewidth}
\center{\includegraphics[width=1\linewidth]{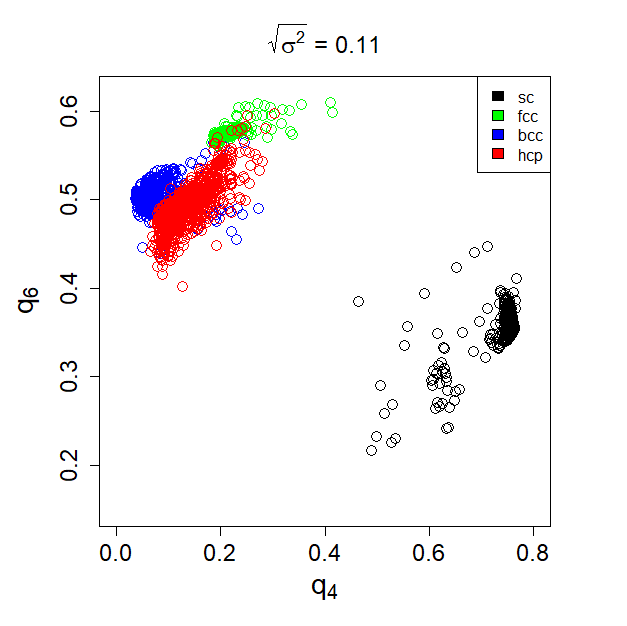} \\ (b)}
\end{minipage}
\vfill
\begin{minipage}[h]{0.40\linewidth}
\center{\includegraphics[width=1\linewidth]{./resources/figures/noise_reduction/points/q48.png} \\ (c)}
\end{minipage}
    \begin{minipage}[h]{0.40\linewidth}
\center{\includegraphics[width=1\linewidth]{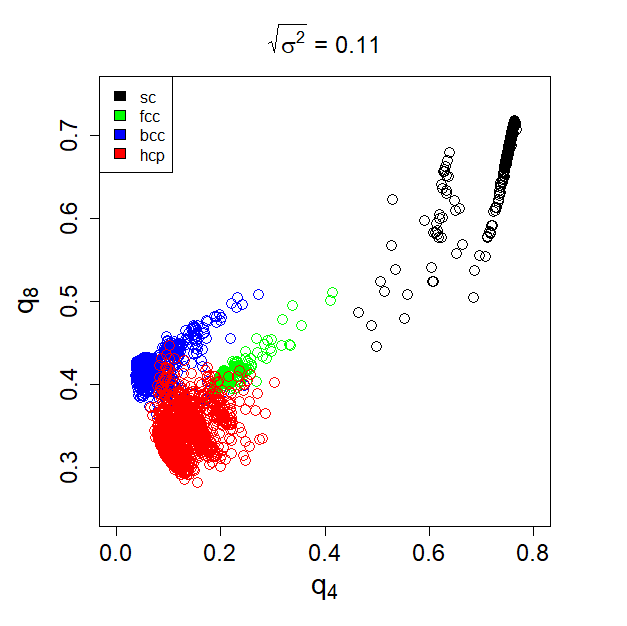} \\ (d)}
\end{minipage}
\vfill
    \begin{minipage}[h]{0.40\linewidth}
\center{\includegraphics[width=1\linewidth]{./resources/figures/noise_reduction/points/q68.png} \\ (e)}
\end{minipage}
    \begin{minipage}[h]{0.40\linewidth}
\center{\includegraphics[width=1\linewidth]{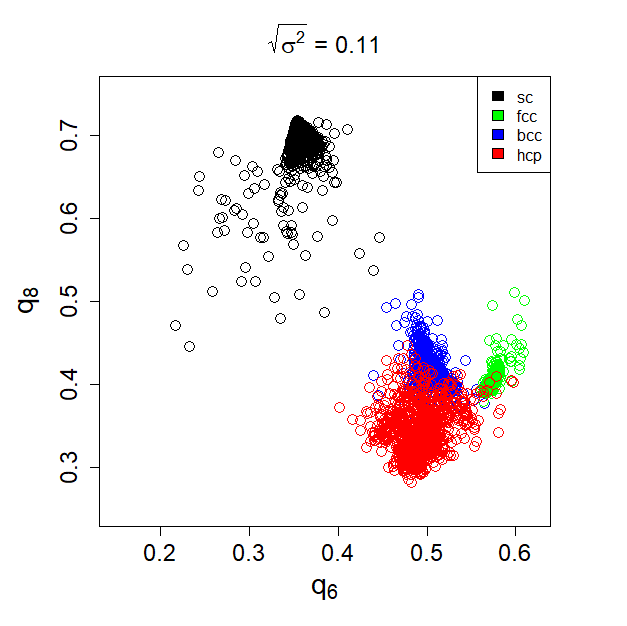} \\ (f)}
\end{minipage}
    \caption{Local bond order parameters after averaging procedure $q_4-q_6$, $q_4-q_8$, $q_6-q_8$ match particles (a, c, e) and cm (b, d, f) for test structures.}
    \label{fig:test_reduce_noise_comp1}
    }
\end{figure}

\begin{figure}[H]
\center{
    \begin{minipage}[h]{0.45\linewidth}
\center{\includegraphics[width=1\linewidth]{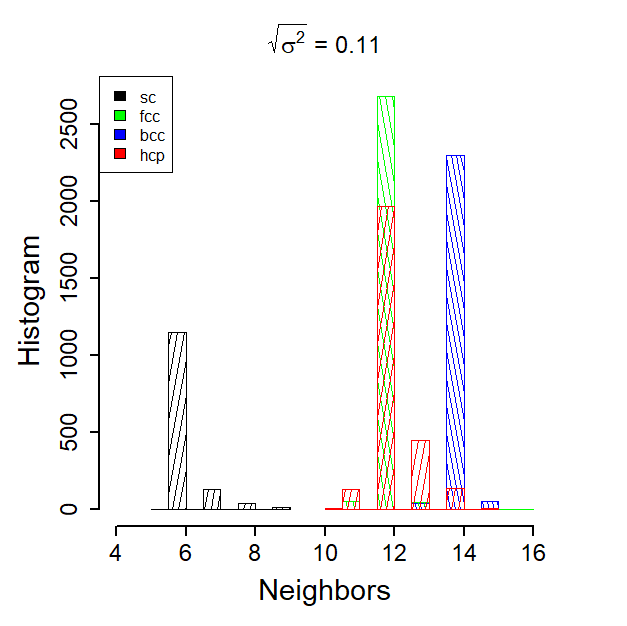} \\ (a)}
\end{minipage}
\begin{minipage}[h]{0.45\linewidth}
\center{\includegraphics[width=1\linewidth]{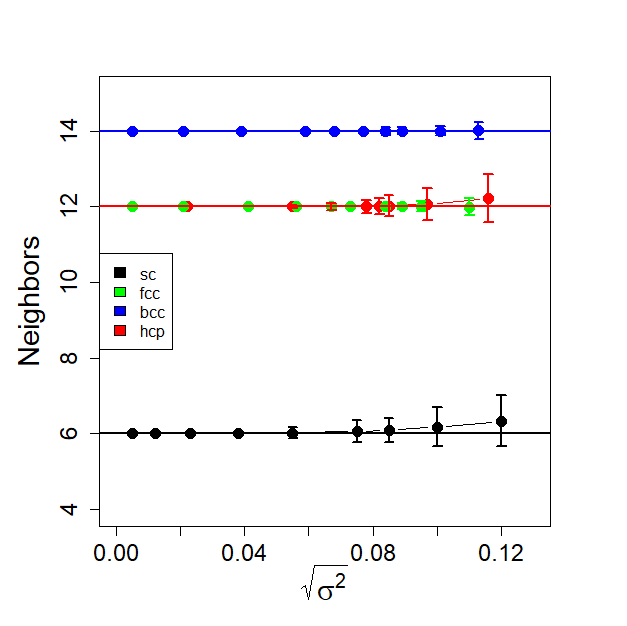} \\ (b)}
\end{minipage}
    \vfill
    \begin{minipage}[h]{0.45\linewidth}
\center{\includegraphics[width=1\linewidth]{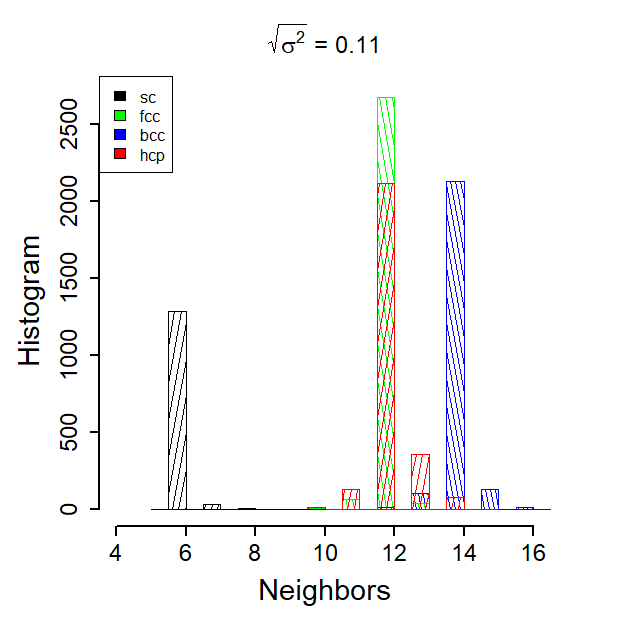} \\ (d)}
\end{minipage}
    \begin{minipage}[h]{0.45\linewidth}
\center{\includegraphics[width=1\linewidth]{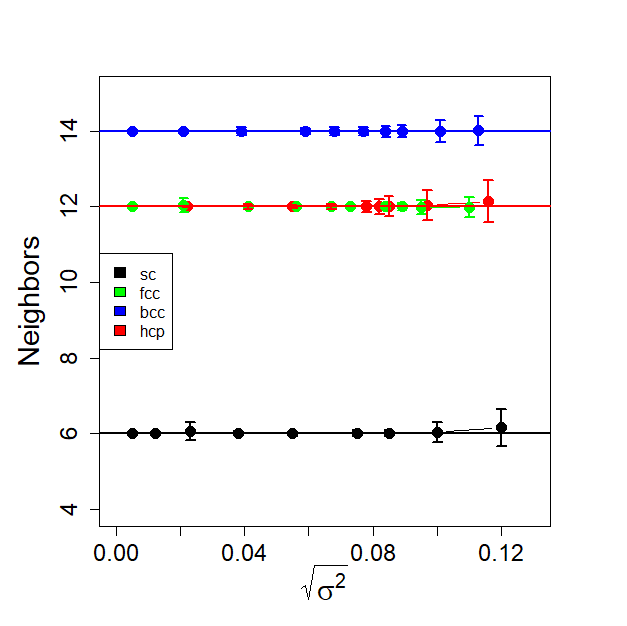} \\ (e)}
\end{minipage}
    \caption{Neighbors after averaging procedure $q_4$, $q_6$, $q_8$ match particles (a, b) and cm (c, d) for test structures.}
    \label{fig:test_reduce_noise_comp4}
    }
\end{figure}

\begin{figure}[H]
\center{
    \begin{minipage}[h]{0.42\linewidth}
\center{\includegraphics[width=1\linewidth]{./resources/figures/noise_reduction/points/q4.png} \\ (a)}
\end{minipage}
\begin{minipage}[h]{0.42\linewidth}
\center{\includegraphics[width=1\linewidth]{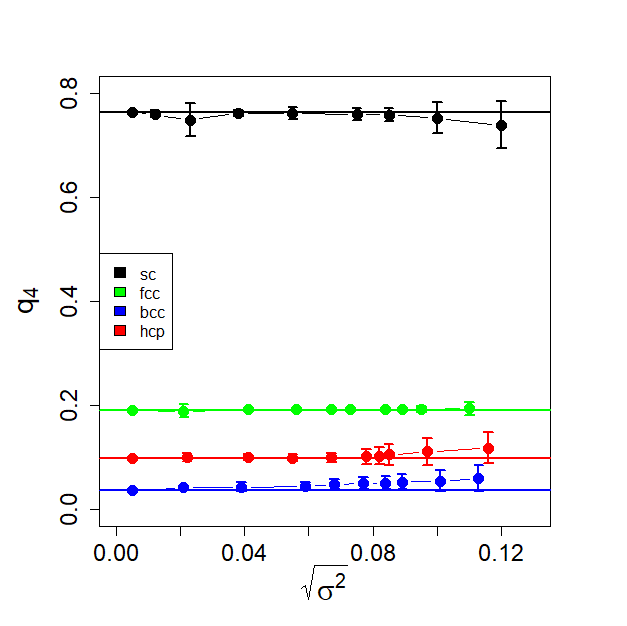} \\ (b)}
\end{minipage}
\vfill
\begin{minipage}[h]{0.42\linewidth}
\center{\includegraphics[width=1\linewidth]{./resources/figures/noise_reduction/points/q6.png} \\ (c)}
\end{minipage}
    \begin{minipage}[h]{0.42\linewidth}
\center{\includegraphics[width=1\linewidth]{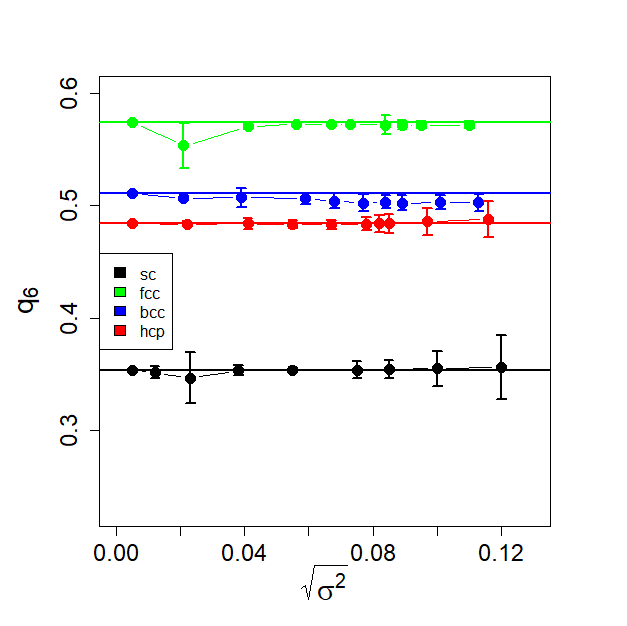} \\ (d)}
\end{minipage}
\vfill
    \begin{minipage}[h]{0.42\linewidth}
\center{\includegraphics[width=1\linewidth]{./resources/figures/noise_reduction/points/q8.png} \\ (e)}
\end{minipage}
    \begin{minipage}[h]{0.42\linewidth}
\center{\includegraphics[width=1\linewidth]{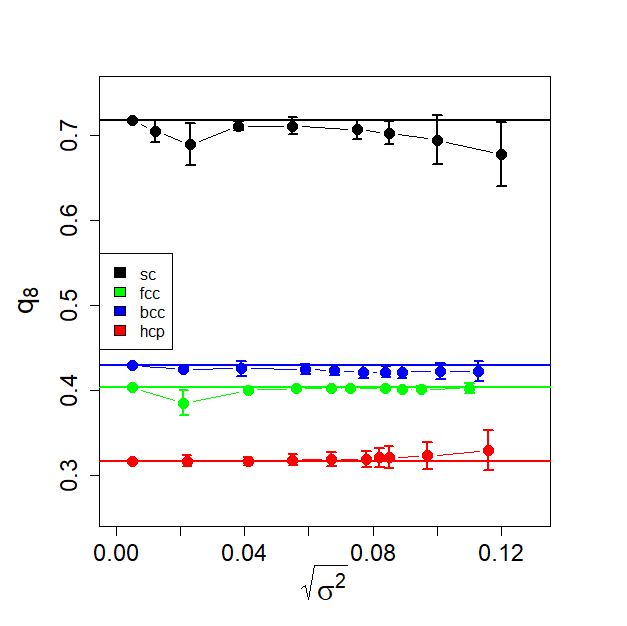} \\ (f)}
\end{minipage}
    \caption{Mean local bond order parameters after averaging procedure $q_4$, $q_6$, $q_8$ match particles (a, b, c) and cm (b, d, f) for test structures.}
    \label{fig:test_reduce_noise_comp2}
    }
\end{figure}

\begin{table}[H]
\center{
\begin{tabular}{|c|c||c|c|c|}
\hline
Melt & High E & $\bar{q}_4$ & $\bar{q}_6$ & $\bar{q}_8$\\
\hline
\hline
1 & -1000 & 0.2232338 & 0.3992973 & 0.3341042 \\
\hline
2 & -2040 & 0.2229070 & 0.4007197 & 0.3331288 \\
\hline
3 & -2400 & 0.2240491 & 0.4015599 & 0.3348154\\
\hline
4 & -3000 & 0.2261278 & 0.4034265 & 0.3354119\\
\hline

\end{tabular}
    \caption{Melt; $\bar{q}_4$, $\bar{q}_6$, $\bar{q}_8$  of the structures found in the system without walls, $L_x=L_y=20$, $L_z=19$.}
\label{tab:found_param1}
}
\end{table}

\begin{table}[H]
\center{
\begin{tabular}{|c|c||c|c|c|}
\hline
Crystal 1 & Low E & $\bar{q}_4$ & $\bar{q}_6$ & $\bar{q}_8$ \\
\hline
\hline
1 & -5727 & 0.1491924 & 0.4910839 & 0.2875898\\
\hline
2 & -5717 & 0.1507784 & 0.4928109 & 0.2845559\\
\hline
3 & -5605 & 0.1515706 & 0.4928678 & 0.2852961\\
\hline

\end{tabular}
    \caption{Crystal 1; $\bar{q}_4$, $\bar{q}_6$, $\bar{q}_8$  of the structures found in the system without walls, $L_x=L_y=20$, $L_z=19$.}
\label{tab:found_param2}
}
\end{table}

\begin{table}[H]
\center{
\begin{tabular}{|c|c||c|c|c|}
\hline
Crystal 2 & Low E & $\bar{q}_4$ & $\bar{q}_6$ & $\bar{q}_8$\\
\hline
\hline
1 & -5727 & 0.2013380 & 0.5557263 & 0.3757146\\
\hline
2 & -5717 & 0.1999767 & 0.5561825 & 0.3760433 \\
\hline
3 & -5605 & 0.2012632 & 0.5559024 & 0.3763463\\
\hline
\end{tabular}
    \caption{Crystal 2; $\bar{q}_4$, $\bar{q}_6$, $\bar{q}_8$  of the structures found in the system without walls, $L_x=L_y=20$, $L_z=19$.}
\label{tab:found_param3}
}
\end{table}

\begin{figure}[h]
\center{
    \begin{minipage}[h]{0.32\linewidth}
\center{\includegraphics[width=1\linewidth]{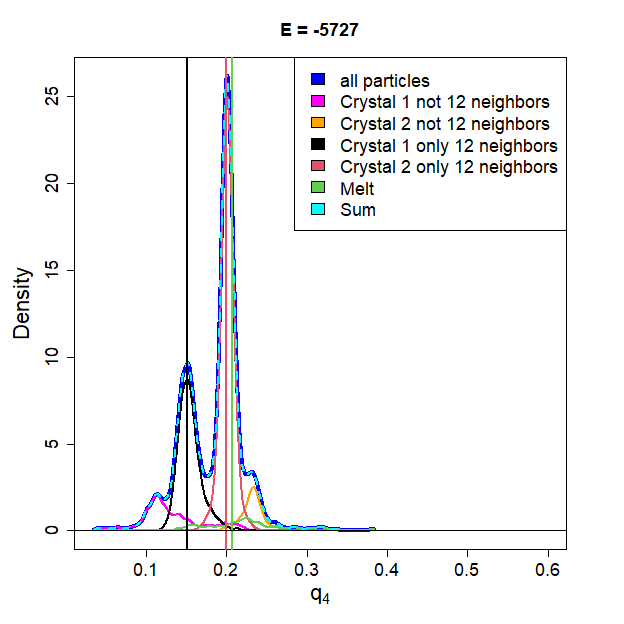} \\ (a)}
\end{minipage}
\begin{minipage}[h]{0.32\linewidth}
\center{\includegraphics[width=1\linewidth]{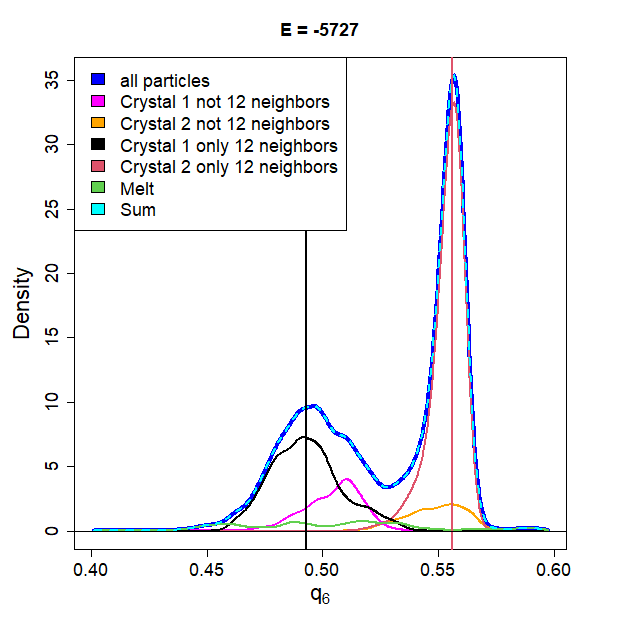} \\ (b)}
\end{minipage}
    \begin{minipage}[h]{0.32\linewidth}
\center{\includegraphics[width=1\linewidth]{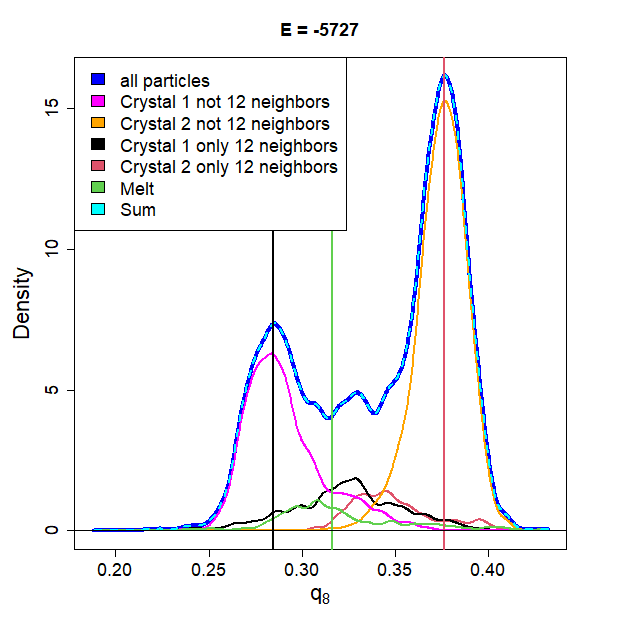} \\ (c)}
\end{minipage}
    \caption{Local bond order parameters distribution for the bulk system $E=-5727$ , $L_x=L_y=20$, $L_z=20$ after averaging coordinates
    \label{fig:separ_dens_supl}
    }
}  
\end{figure}

\end{document}